\documentclass[journal]{IEEEtran}

\usepackage{epsfig,amsmath,amssymb,amsbsy,amsthm,epsf,amsthm,scalefnt,subfig,multirow,algorithm,algorithmic,mathtools}
\usepackage{xcolor}
\usepackage{float}
\usepackage{cite}
\usepackage[T1]{fontenc}

\def\inf{\mathop{\mathrm{inf}}}

\def\b0{{\pmb{0}}} 

\newcommand\ignore[1]{}

\begin{document}

\title{Secure Distributed Dynamic State Estimation in Wide-Area Smart Grids}

\author{Mehmet Necip~Kurt,~
        Yasin~Y{\i}lmaz,~
        and~~Xiaodong~Wang
\thanks{This work was supported in part by the U.S. National
Science Foundation (NSF) under Grant ECCS-1405327, and in part by the
U.S. Office of Naval Research under Grant N000141410667. The work of
Y. Y{\i}lmaz was supported in part by the NSF under Grant CNS-1737598.}%
\thanks{M. N. Kurt and X. Wang are with the Department
of Electrical Engineering, Columbia University, New York, NY 10027, USA (e-mail: m.n.kurt@columbia.edu; wangx@ee.columbia.edu).}
\thanks{Y. Y{\i}lmaz is with the Department of Electrical Engineering, University of South Florida, Tampa, FL 33620, USA (e-mail: yasiny@usf.edu).}\vspace{-2.2ex}}

\maketitle

\begin{abstract}
Smart grid is a large complex network with a myriad of vulnerabilities, usually operated in adversarial settings and regulated based on estimated system states. In this study, we propose a novel highly secure distributed dynamic state estimation mechanism for wide-area {(multi-area)} smart grids, composed of geographically separated subregions, each supervised by a local control center. We firstly propose a distributed state estimator assuming regular system operation, that achieves near-optimal performance based on the local Kalman filters and with the exchange of necessary information between local centers. To enhance the security, we further propose to (i) protect the network database and the network communication channels against attacks and data manipulations via a blockchain (BC)-based system design, where the BC operates on the peer-to-peer network of local centers, (ii) locally detect the measurement anomalies in real-time to eliminate their effects on the state estimation process, and (iii) detect misbehaving (hacked/faulty) local centers in real-time via a distributed trust management scheme over the network. We provide theoretical guarantees regarding the false alarm rates of the proposed detection schemes, where the false alarms can be easily controlled. Numerical studies illustrate that the proposed mechanism offers reliable state estimation under regular system operation, timely and accurate detection of anomalies, and good state recovery performance in case of anomalies.
\end{abstract}

\begin{keywords}
\noindent Smart grid, distributed, secure, dynamic state estimation, Kalman filter, blockchain, real-time anomaly detection, trust management.
\end{keywords}

\section{Introduction} \label{sec:intro}

\begin{figure}[t]
\center
  \includegraphics[width=84mm]{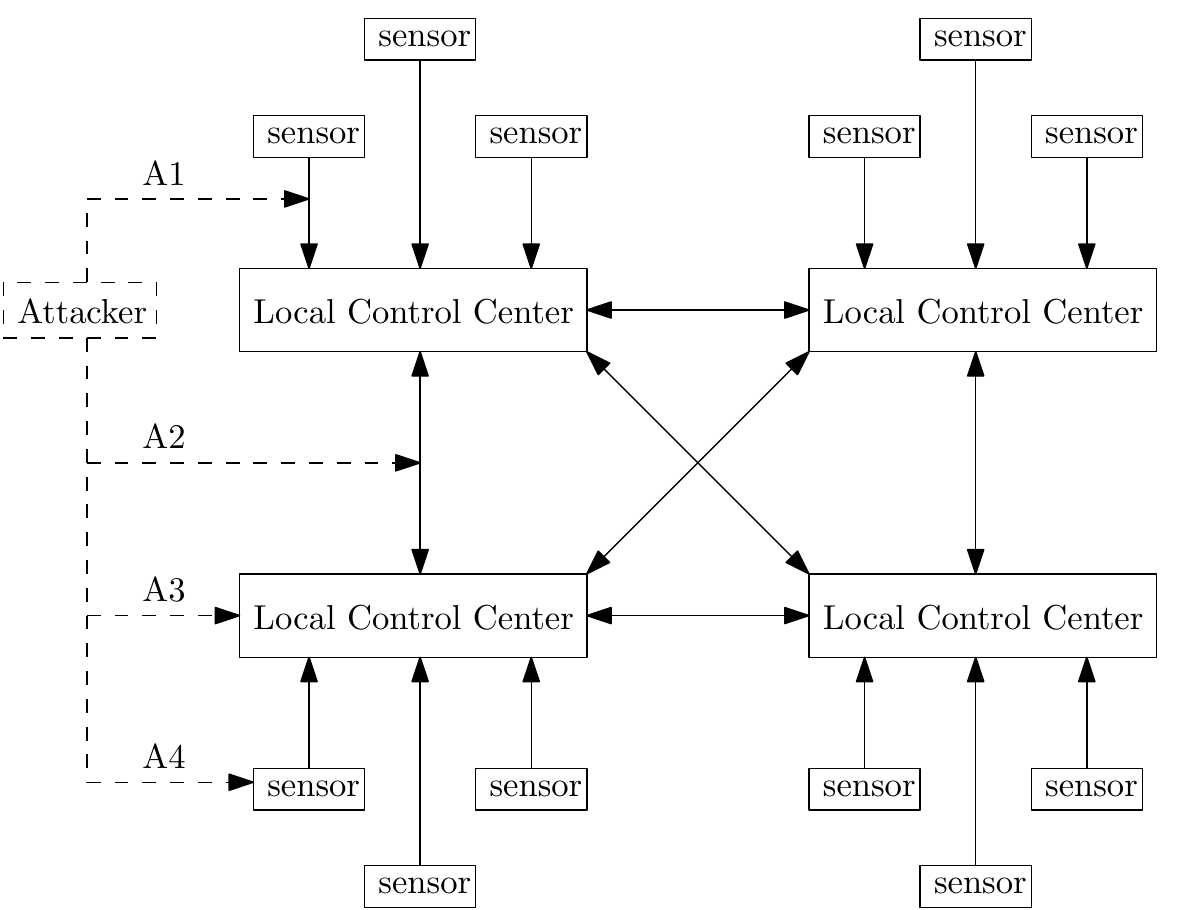}
\caption{\footnotesize An illustration of the vulnerabilities of wide-area smart grids against malicious adversaries. The solid lines illustrate the network communication channels and the dashed lines illustrate possible attacks. The attacker can (i) manipulate, jam or block the communication channels (A1, A2), (ii) hack a local control center to manipulate its operation or to tamper with the network database (A3), and (iii) hack a {sensor} to corrupt its online measurements (A4).}
\label{fig:attacks}
\end{figure}

The next-generation electrical power grid, i.e., the smart grid, is vulnerable to a variety of cyber/physical system faults and hostile cyber threats \cite{Liang16,lopes2016vulnerabilities,grochocki2012ami,Mylrea17}. In particular, random anomalies such as node and topology failures might occur all over the network. Moreover, malicious attackers can deliberately manipulate the network operation and tamper with the network data, from sources such as {sensors}, control centers, network database, and network communication channels (see Fig.~\ref{fig:attacks}). False data injection (FDI), jamming, and denial of service (DoS) attacks are well-known attack types \cite{lopes2016vulnerabilities,Necip18,liu2011false,Necip19,Necip19a}. Further, Internet of Things (IoT) botnets can be used to target against critical infrastructures such as the smart grid \cite{Yilmaz17IoT,soltan18blackiot,bertino2017botnets}.

In the smart grid, online state estimates are utilized to make timely decisions in critical tasks such as load-frequency control and economic dispatch \cite{Ashok17}. Hence, a fundamental task in the grid is reliable state estimation based on online measurements. On the other hand, the main objective of the adversaries is to damage/mislead the state estimation mechanism in order to cause wrong/manipulated decisions, resulting in power blackouts or manipulated electricity prices \cite{Xie10}. Additionally, random system faults may degrade the state estimation performance. Our objective in this study is to design a highly secure and resilient state estimation mechanism for wide-area smart grids, that provides reliable state estimates in a fully-distributed manner, even in the case of cyber-attacks and other network anomalies.

\subsection{Background and Related Work} \label{sec:motivation}

\subsubsection{Secure Dynamic State Estimation}

Feasibility of dynamic modeling and efficiency of dynamic state estimation have been widely discussed and various dynamic models have been proposed for the power grids \cite{Debs70,Jain08,yang2016false,da1983,huang2007}. The general consensus is that the dynamic modeling better captures the time-varying characteristics of the power grid and dynamic state estimators are more effective to track the system state compared to the conventional static least squares (LS) estimators. Moreover, state forecasting capability achieved with dynamic estimators is quite useful in real-time operation and security of the grid \cite{Debs70,Jain08,Necip18}.

In the literature, various techniques have been proposed to make the dynamic state estimation mechanism secure/robust against outliers, heavy-tailed noise processes, model uncertainties including unknown noise statistics, rank-deficient observation models, and cyber-attacks, etc., \cite{Gandhi10,chang2017unified,zhang2005robust,zhao2017robust,Huang17,Agamennoni12,HWang17,chang2018secure,mo2016secure,Fawzi14,graham2015robust,Shoukry17,wang2018robust,da1987efficient,mu2014novel,Taha18}. For instance, robust statistics-based approaches \cite{Gandhi10,chang2017unified,zhang2005robust,zhao2017robust} aim to suppress the effects of outliers by assigning less weights to more significant outliers. As the outliers are still incorporated into the state estimation process, their effects are not completely eliminated, and due to the recursive nature of dynamic state estimators, errors are accumulated over time and the corresponding state estimator breaks down at some point, i.e., fails to keep track of the system operating point. The estimator also breaks down in case of gross outliers. Furthermore, this approach is based on the solution of an iterative weighted LS problem, repeated at each time, that might be prohibitive for real-time processing.

To deal with outliers, another method is modeling the system noise as a heavy-tailed, e.g., Student's t or Laplace, distribution \cite{Huang17,Agamennoni12,HWang17}. This method can handle (nominal) outliers observed during the regular system operation, however, it is expected to be ineffective against attacks and faults that behave significantly different from nominal outliers. Further, some studies are based on the assumption that attacks are sparse over the network or with bounded magnitudes \cite{chang2018secure,mo2016secure,Fawzi14,graham2015robust,Shoukry17}, which significantly limits the robustness of the corresponding state estimators. This is because imposing restrictions on the capabilities or strategies of attackers make the corresponding state estimation mechanism robust against only a certain subset of attacks. Yet another approach is to completely ignore the outliers in the state estimation process \cite{wang2018robust,da1987efficient,mu2014novel,Valverde11}. In the literature, this approach is usually based on the sample-by-sample classification of observations as either outlier or non-outlier. Although it might be useful to detect and eliminate the effects of gross outliers, the corresponding state estimator is expected to fail if small/moderate-level (difficult-to-detect) outliers are observed in a persistent manner. {Moreover, it might fail even in the nominal case due to the rejection of nominal outliers. Finally, several dynamic state estimation techniques are evaluated in \cite{Taha18} for the nonlinear power system model and it is shown that the cubature Kalman filter is more robust to model uncertainties and attacks/anomalies compared to the extended Kalman filter and the unscented Kalman filter.}

Distributed dynamic state estimation has been also studied extensively, see e.g., \cite{Khalid13} for a review of the distributed Kalman filtering techniques. Particularly, two main architectures are considered: hierarchical and fully-distributed. In the former, there exists a central controller coordinating multiple local controllers, while in the latter, no central controller exists. In the hierarchical schemes, the information filter, an algebraic equivalence to the centralized Kalman filter, can be useful to fuse the information processed across the local controllers in a simple manner \cite{Mutambara98,TSIPN18}. The fully-distributed schemes usually require an iterative consensus mechanism to reduce the disagreement of the local controllers on the estimates of common state variables \cite{Khalid13,Olfati-Saber07}. More recently, a distributed extended Kalman filter is proposed in \cite{Rostami18} based on the internodal transformation theory, not requiring iterative data exchanges. A recent survey of model-based distributed filtering and control including the distributed Kalman filtering and secure control algorithms for large-scale cyber-physical systems is presented in \cite{Ding19}. For instance, \cite{Ding18} studies the secure control of stochastic linear systems against deception attacks, that manipulate sensor measurements or control inputs.

\subsubsection{Secure Distributed System Design}

Traditionally, through the supervisory control and data acquisition (SCADA) system, the power grid is controlled in a centralized manner. In particular, the system-wide data are collected, stored, and processed at a single node. Considering the increasing speed and size of the measurement data, collecting and processing such large volume of data in real-time at a single center seem practically infeasible in the modern power grids \cite{Rostami18}. Moreover, the traditional implementation is based on the assumption that the centralized node is completely trustable. In practice, however, the centralized node can be the weakest point of the network in terms of security. This is because by hacking only the centralized node, adversaries can arbitrarily modify the control decisions and the network database. On the contrary, hacking a distributed system is usually more difficult for attackers, especially for the smart grid that is distributed over a geographically wide region. Therefore, distributing the computation and the trust over the network can be useful to achieve a more feasible and a more secure grid.

Blockchain (BC) is an emerging secure distributed database technology, operating on a peer-to-peer (P2P) network with the following key components \cite{puthal2018,nguyen2018survey,Dinh18}: (i) a distributed ledger (chronologically ordered sequence of blocks) that is shared and synchronized over the network where each block is cryptographically linked to the previous blocks and the ledger is resistant/immune to modifications, (ii) advanced cryptography enabling secure data exchanges and secure data storage, and (iii) a mutual consensus mechanism, that enables collective verification/validation on the integrity of exchanged and stored data, and thereby distributes trust over the network instead of relying on a single node/entity. BC technology was firstly used in financial applications \cite{Bitcoin,Ethereum} but due to its security-by-design and the distributed nature without needing any trusted third party, it has been applied to diverse fields such as vehicular networks, supply chain management, cognitive radio, and insurance \cite{tapscott2016,Kotobi18,ZYang18}.

The smart grid critically relies on a database and network communication channels, both are quite vulnerable to attacks and manipulations. As a countermeasure, an effective approach is the detection and then mitigation of such threats. On the other hand, a significantly better approach is the prevention of the threats as much as possible. In this direction, the BC technology has a great potential due to its advanced data protection/attack prevention capabilities. Hence, we aim to integrate some salient features of the BC technology to the smart grid in order to improve the resilience of the system, particularly to protect the system database and the communication channels.

Research on the integration of BC to smart grids has mainly focused on secure energy transactions/trade so far \cite{Aitzhan18,lombardi2018,Exergy,Mylrea17}, and only a few studies have examined the integration of BC to make the power supply system more secure, see \cite{GLiang18} and \cite{zhaoyang2018}, where in both studies, the grid is protected by securely storing all the system-wide measurements at every meter. However, this seems infeasible in many aspects, e.g., meters are small-size devices with limited memory, power, and processing capabilities \cite{lombardi2018} and hence not suitable to perform advanced computations such as encryption/decryption, to store the distributed ledger, and to constantly communicate with all other meters in the network. Moreover, since the measurements are collected mainly to estimate the system state, the BC-based system can be designed to protect the state estimation mechanism in a more direct way, rather than to protect the entire history of raw measurement data, that then enables secure state estimation.

Although BC can be useful to secure the network database and the communication channels, the online {sensor} measurements are still vulnerable to attacks and faults. We would like to design a state estimation mechanism that is secure against all types of anomalies. Towards this goal, as a complement to the BC-based data protection, robust bad data detection and mitigation, i.e., state recovery, schemes need to be integrated into the state estimation mechanism. We have recently proposed in \cite{Necip19} a robust dynamic state estimation scheme for the smart grid, in case the attack models are known (with some unknown parameters). In practice, unknown attacks/anomalies may occur in the smart grid as it has many vulnerabilities and attackers might have arbitrary strategies. Hence, in general, anomaly/attack models need to be assumed unknown and the state estimation mechanism should be designed accordingly. Furthermore, in a BC-based distributed system, there is no centralized trusted node to check and recover a node that is faulty or hacked by a malicious entity. Hence, a distributed trust management mechanism needs to be employed over the network to evaluate the trustability of each node against the possibility of misbehaving nodes.

\subsection{Contributions} \label{sec:contributions}

In this study, we propose a novel BC-based resilient system design to achieve secure distributed dynamic state estimation in wide-area smart grids. Our aim is to reduce the risks at each part of this highly complex network, specifically, the network database, {sensors}, local control centers, and network communication channels (see Fig.~\ref{fig:attacks}).

Firstly, assuming regular system operation (no anomaly), we propose a {novel} fully-distributed dynamic state estimation scheme that achieves near-optimal performance thanks to the local Kalman filters and with the exchange of necessary information between local centers. Then, to improve the resilience of the proposed mechanism, we further propose to
\begin{itemize}
  \item use salient features of the emerging BC technology to secure both the network database and the network communication channels against attacks and manipulations,
  \item embed {novel} online anomaly detection schemes into the state estimation mechanism to make it secure against measurement anomalies, and
  \item detect and eliminate the effects of misbehaving nodes in real-time via a novel distributed trust management mechanism over the network.
\end{itemize}

We provide theoretical guarantees regarding the false alarm rates of the proposed online detection schemes, where the false alarms can be easily controlled by the system designer. {Finally, since the proposed anomaly detection, trust management, and state recovery schemes do not assume any knowledge about the type/cause of the anomaly/misbehavior, they are all robust to unknown attacks/anomalies.}

\subsection{Organization and Notations}

The remainder of the paper is organized as follows. Sec.~\ref{sec:system} presents the system model. Sec.~\ref{sec:blockchain} describes the proposed BC-based secure system design. Sec.~\ref{sec:SE-nominal} discusses the proposed distributed state estimation mechanism under regular (non-anomalous) network operation. Sec.~\ref{sec:robust} explains the proposed online anomaly detection scheme against measurement anomalies and the corresponding state recovery scheme. Sec.~\ref{sec:trust} discusses the proposed distributed trust management scheme against misbehaving nodes. Sec.~\ref{sec:summary} then summarizes the proposed mechanism. Sec.~\ref{sec:numerical} illustrates the advantages of the proposed mechanism over a simulation setup. Finally, Sec.~\ref{sec:conc} concludes the paper.

\textit{Notations:}
Boldface letters denote vectors and matrices. $\mathbb{R}$ denotes the set of real numbers. $\mathcal{N}(\pmb{\mu},\pmb{\Sigma})$ denotes the Gaussian probability density function (pdf) with mean $\pmb{\mu}$ and covariance matrix $\pmb{\Sigma}$. $\mathbf{I}_m$ denotes an ${m \times m}$ identity matrix. $\mathrm{P}(\cdot)$ and $\mathrm{E}(\cdot)$ denote the probability and the expectation operators, respectively. $\mathbb{I}(\cdot)$ denotes the indicator function. $\log(\cdot)$ denotes the natural logarithm and $e$ denotes the Euler's number. $|\cdot|$ denotes the cardinality of a set. $\emptyset$ denotes an empty set. $\mathcal{S}_1 \backslash \mathcal{S}_2$ denotes the set of elements belonging to $\mathcal{S}_1$ but not belonging to $\mathcal{S}_2$. $\inf$, $\sup$, and $\max$ denote the infimum, supremum, and maximum operators, respectively, and $(\cdot)^+ \triangleq \max\{\cdot,0\}$. Finally, ${\cdot}^{\scriptscriptstyle \mathrm{T}}$ denotes the transpose operator.

\section{System Model} \label{sec:system}

We consider a smart power grid with $N$ buses and $K$ sensors. System state at time $t$, $\mathbf{x}_t = [x_{1,t}, \dots, x_{{2 N - 1},t}]^{\scriptscriptstyle \mathrm{T}}$, represents voltage {magnitudes and} phase angles of the buses, where a bus is chosen as the reference {for the phase angles. The measurements are real and reactive power flow, power injection, and voltage magnitude measurements.} The measurement taken at {sensor} $k \in \{1, \dots, K\}$ at time $t$ is denoted with $y_{k,t}$ and the measurement vector is denoted with $\mathbf{y}_t = [y_{1,t}, \dots, y_{K,t}]^{\scriptscriptstyle \mathrm{T}}$. We model the grid as a discrete-time linear dynamic system as follows \cite{Debs70,Jain08,da1987efficient,Nishiya82,Abur04}:
\begin{gather} \label{eq:state_upd}
\mathbf{x}_t = \mathbf{A} \mathbf{x}_{t-1} + \mathbf{v}_t, \\ \label{eq:meas_model}
\mathbf{y}_t = \mathbf{H} \mathbf{x}_t + \mathbf{w}_t,
\end{gather}
where $\mathbf{A} \in \mathbb{R}^{{2 N - 1} \times {2 N - 1}}$ is the state transition matrix, $\mathbf{H} \in \mathbb{R}^{K \times {2 N - 1}}$ is the measurement matrix, $\mathbf{v}_t = [v_{1,t}, \dots, v_{{2 N - 1},t}]^{\scriptscriptstyle \mathrm{T}}$ is the process noise vector, and ${\mathbf{w}_t = [w_{1,t}, \dots, w_{K,t}]^{\scriptscriptstyle \mathrm{T}}}$ is the measurement noise vector. We assume that $\mathbf{v}_t$ and $\mathbf{w}_t$ are independent AWGN processes where ${\mathbf{v}_t \sim \mathbf{\mathcal{N}}(\mathbf{0},\sigma_v^2 \, \mathbf{I}_{{2 N - 1}})}$ and ${\mathbf{w}_t \sim \mathbf{\mathcal{N}}(\mathbf{0},\sigma_w^2 \, \mathbf{I}_K)}$.

{\subsection{State-Space Model}}

To model the time behavior of the power system state, several attempts have been made. Firstly, considering small time intervals between consecutive measurements, a linear state transition model is assumed \cite{da1987efficient,Jain08}. Further, a quasi-static state model is commonly employed where the system state is periodic over a day since the system loads typically have a daily cycle \cite{da1987efficient,Debs70}. Moreover, a linear exponential smoothing model is proposed in \cite{da1983}, where the effects of past measurements on the state estimates are reduced over time. In the simplest case, the system state is assumed to be the same with the previous time step, up to a random uncertainty \cite{Debs70}, i.e.,
\begin{equation}\nonumber
\mathbf{x}_t = \mathbf{x}_{t-1} + \mathbf{v}_t,
\end{equation}
where the process noise $\mathbf{v}_t$ corresponds to the random uncertainty. In a more general form, we can write
\begin{equation}\nonumber
\mathbf{x}_t = \mathbf{A} \mathbf{x}_{t-1} + \mathbf{v}_t.
\end{equation}

In the actual power system operation, the relationship between the measurements and the system state is nonlinear. Hence, we have
\begin{equation}\nonumber
\mathbf{y}_t = h(\mathbf{x}_t) + \mathbf{w}_t,
\end{equation}
where $h(\cdot)$ is a nonlinear function. Linearizing the system around an operating point $\mathbf{x}_0$, we obtain
\begin{equation}\nonumber
\mathbf{y}_t = \mathbf{H} \mathbf{x}_t + \mathbf{w}_t,
\end{equation}
where
\begin{equation}\nonumber
\mathbf{H} = \frac{\partial h(\mathbf{x})}{\partial \mathbf{x}}{\bigg |}_{\mathbf{x} = \mathbf{x}_0}
\end{equation}
is the Jacobian matrix.

The parameters of the state-space model, particularly the state transition matrix and the process and measurement noise variances can be determined from power system data obtained under regular operating conditions \cite{Debs70,Jain08}.

{\subsection{Wide-Area Monitoring Model}}

The wide-area smart grid is composed of geographically separated subregions (see Fig.~\ref{fig:ieee14bus}). Each subregion contains a set of {sensors}, supervised by a local (control) center. Since the {sensors} are distributed over the network, each local center partially observes the measurement vector $\mathbf{y}_t$. Assuming that the grid is composed of $L$ subregions and the subset of {sensors} in the $\ell$th subregion is denoted by $\mathcal{R}^{\ell}$, the measurement vector $\mathbf{y}_t$ is decoupled into $L$ sub-vectors $\mathbf{y}_t^{\ell} \in \mathbb{R}^{K^{\ell}}$, $\ell = 1, \dots, L$, where $\mathbf{y}_t^{\ell} \triangleq \{y_{k,t} \,|\, k \in \mathcal{R}^{\ell}\}$ denotes (with an abuse of notation) the measurement vector of the $\ell$th local center at time $t$ and $K^{\ell} \triangleq |\mathcal{R}^{\ell}|$ is the number of {sensors} in the $\ell$th subregion. Since each {sensor} belongs to only one subregion, for any two subregions $\ell$ and $j$, $\mathbf{y}_t^{\ell}$ and $\mathbf{y}_t^{j}$ do not overlap and we have $\sum_{\ell=1}^{L} K^{\ell} = K$.

The smart grid is an interconnected system, where there exist tie-lines between neighboring subregions (see Fig.~\ref{fig:ieee14bus}) that leads to some common (shared) state variables between neighboring local centers. Hence, denoting the state vector of the $\ell$th local center at time $t$ by $\mathbf{x}_t^{\ell} \in \mathbb{R}^{N^{\ell}}$, for any two neighboring local centers $\ell$ and $j$, $\mathbf{x}_{t}^{\ell}$ and $\mathbf{x}_{t}^{j}$ might overlap. This implies $\sum_{\ell=1}^{L} {N^{\ell}} \geq {2 N - 1}$. In general, if the state transition matrix $\mathbf{A}$ is non-diagonal, additional state variables might be shared between neighboring or non-neighboring local centers due to dependencies between state variables over time through the state transition matrix. In this study, for the simplicity of the presentation, we assume $\mathbf{A}$ is diagonal. Under this assumption, we next determine the state vector of a local center, say the $\ell$th one. {For the derivation of local state vectors in the case of non-diagonal $\mathbf{A}$, please see \cite[Sec.~IV-A]{Necip18}.} Note that in the non-diagonal case, the proposed system design directly extends, where the only difference is that the size of the local state vectors might be larger.

Let $\mathbf{h}_k^{\scriptscriptstyle \mathrm{T}} = [h_{k,1}, \dots, h_{k,{2 N - 1}}]$ be the $k$th row of the measurement matrix $\mathbf{H}$, i.e., ${\mathbf{H}^{\scriptscriptstyle \mathrm{T}} = [\mathbf{h}_1, \dots, \mathbf{h}_K]}$. Then, using \eqref{eq:meas_model}, each measurement $y_{k,t}, k \in \{1,\dots,K\}$ can be written as follows:
\begin{gather} \label{eq:yk_depen}
y_{k,t} = \mathbf{h}_k^{\scriptscriptstyle \mathrm{T}} \mathbf{x}_t + w_{k,t}.
\end{gather}
Based on \eqref{eq:yk_depen}, we can argue that $y_{k,t}$ depends on, equivalently bears information about, the following state variables:
\begin{gather}\nonumber
\mathcal{X}_{y_{k}} \triangleq \{x_{n,t} \,|\, h_{k,n} \neq 0, n = 1,\dots, {2 N - 1}\}.
\end{gather}
Then, the local state vector $\mathbf{x}_{t}^{\ell}$ consists of the union of all such state variables for all the {sensors} $k$ in the subregion $\ell$:
\begin{gather}\nonumber
\mathbf{x}_t^{\ell} = \bigcup_{k \, \in \, \mathcal{R}^{\ell}} \mathcal{X}_{y_{k}}.
\end{gather}

For each local center $\ell \in \{1, \dots, L\}$, we then have the following local state transition model:
\begin{gather} \label{eq:local_states}
\mathbf{x}_{t}^{\ell} = \mathbf{A}^{\ell} \mathbf{x}_{t-1}^{\ell} + \mathbf{v}_t^{\ell},
\end{gather}
and the following local measurement model:
\begin{gather} \label{eq:local_monitoring}
\mathbf{y}_t^{\ell} = \mathbf{H}^{\ell} \mathbf{x}_{t}^{\ell} + \mathbf{w}_t^{\ell},
\end{gather}
where the local state transition matrix, $\mathbf{A}^{\ell} \in \mathbb{R}^{N^{\ell} \times N^{\ell}}$, and the local measurement matrix, $\mathbf{H}^{\ell} \in \mathbb{R}^{K^{\ell} \times N^{\ell}}$, can be easily obtained from $\mathbf{A}$ and $\mathbf{H}$, respectively. Moreover, the local process noise vector, $\mathbf{v}_t^{\ell} \in \mathbb{R}^{N^{\ell}}$, is a sub-vector of $\mathbf{v}_t$ corresponding to $\mathbf{x}_{t}^{\ell}$. Similarly, the local measurement noise vector, $\mathbf{w}_t^{\ell} \in \mathbb{R}^{K^{\ell}}$, is a sub-vector of $\mathbf{w}_t$ corresponding to $\mathbf{y}_{t}^{\ell}$. \\

\section{Blockchain-Based Secure System Design} \label{sec:blockchain}

\subsection{Overview of the Proposed System}

We consider a distributed P2P network of local centers where each node (local center) can communicate with all other nodes (see Fig.~\ref{fig:attacks}). We aim to design a system in which the nodes collaborate with each other to perform the state estimation task in a safe and reliable manner. For a reliable distributed dynamic state estimation, particularly the Kalman filter, we need safe updates and hence the following three items must be secure/reliable at each time $t$:
\begin{itemize}
  \item state estimates of the previous time $t-1$,
  \item {sensor} measurements acquired at the current time $t$, and
  \item the nodes functioning in the state estimation process, i.e., the local centers.
\end{itemize}
In other words, at each time, we need to make sure that the previous state estimates are not modified, the online {sensor} measurements are not anomalous, and the nodes are working according to predesigned network rules. Furthermore, in case of an anomaly over the network, the state estimates can be recovered using the previous reliable state estimates, that requires us to also protect the previous state estimates against tampering. Considering these requirements, our proposed system is composed of the following three main components:
\begin{itemize}
  \item \textit{BC-based data protection/attack prevention:}
BC enhances the security of the grid by reducing the risk of manipulations at the network database and the network communication channels. In particular, to protect the previous state estimates against tampering and make them widely available and accessible over the network against the possibility of node failures and hacking, we record them in a shared distributed ledger that is resistant to alterations. Moreover, we secure the inter-node data exchanges via cryptography against attacks and manipulations.
  \item \textit{Secure state estimation against measurement anomalies:}
Each local center quickly and reliably detects local measurement anomalies and then employs an online state recovery mechanism.
  \item \textit{Distributed trust management:}
All nodes collectively (via voting-consensus) evaluate the trustability of each node, specifically whether the local state estimates provided by a node exhibit an anomalous pattern over time.
\end{itemize}
Firstly, the following subsection explains how we use the BC technology to enhance the security of the state estimation mechanism.

\subsection{Blockchain Mechanism}

The BC operates on the P2P network of local centers. Since each node is pre-specified and pre-authenticated, we have a permissioned (private) BC mechanism \cite{lombardi2018,GLiang18}. In BC-based systems, duties of each node and interactions between nodes are determined via a smart contract, which is a software code that specifies the predefined rules of network operation. In our proposed mechanism, each node collects and analyzes the {sensor} measurements in its subregion, estimates its local state vector, exchanges information with other nodes, performs encryption/decyrption, participates in voting-consensus procedures, and stores the distributed ledger in its memory. The details regarding the duties of the nodes will be more clear in the subsequent sections. Next, we explain the proposed BC mechanism in more detail.

\subsubsection{Data Exchanges} \label{sec:bc-data-exchange}
\begin{figure}[t]
\center
  \includegraphics[width=80mm]{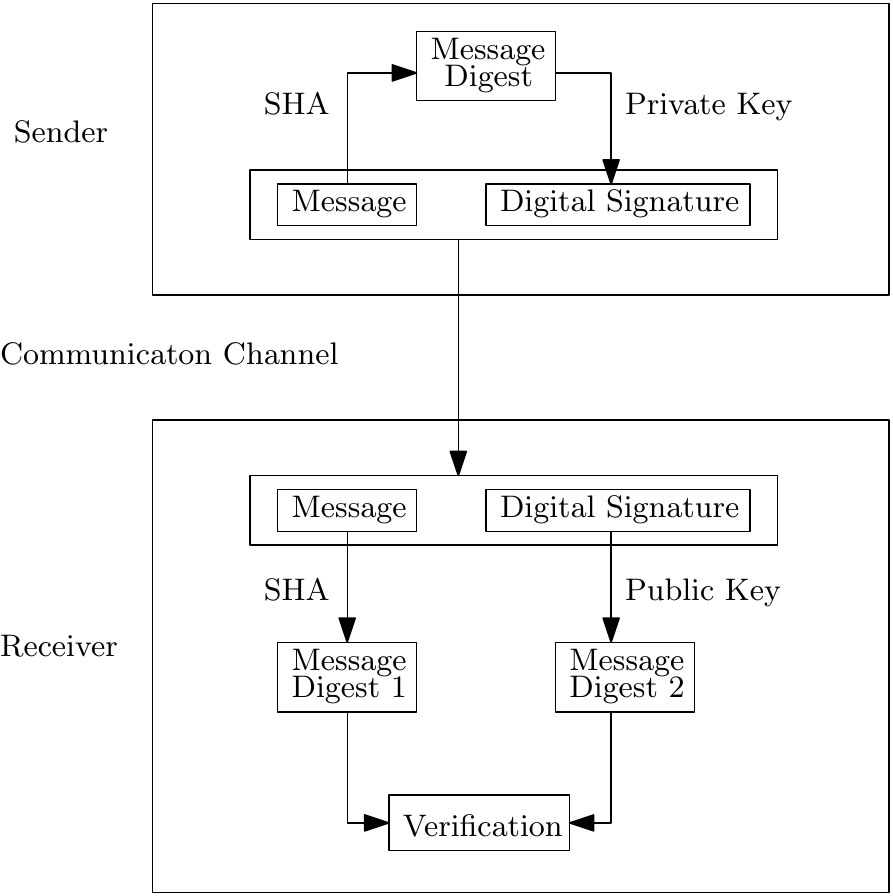}
\caption{\footnotesize Inter-node data exchanges based on the asymmetric encryption mechanism.}
\label{fig:data_exchange}
\end{figure}
In all inter-node data exchanges, we use asymmetric encryption based on the public key infrastructure (see Fig.~\ref{fig:data_exchange}). In this mechanism, each node owns a public-private key pair that forms the digital identity of the node. The public key is available at every node. On the other hand, the private key is available only at its owner. Moreover, a secure hash algorithm (SHA), e.g., SHA-$256$, SHA-$512$, etc. \cite{Pilkington16,Meng18}, is used in the data encryption process. Particularly, in every data exchange, the sender node firstly processes its message via the SHA and obtains the message digest. It then encrypts the message digest via its private key using a signature algorithm, e.g., Elliptic Curve Digital Signature Algorithm \cite{nguyen2018survey,Aitzhan18}, and obtains the digital signature. Finally, it transmits the data package consisting of the message and the corresponding digital signature. The receiver node then decrypts the received digital signature via the public key of the sender node and obtains a message digest. Moreover, it processes the received message via the SHA and obtains another message digest. Only if these two message digests exactly match, the integrity of the received message is verified.

In this procedure, the SHA provides security since it is computationally intractable to obtain the same message digest from two different messages \cite{Pilkington16,Meng18}. The SHA is a one-way function that outputs a fixed-length message digest for an arbitrary-size input message. Let $H(\cdot)$ denote the SHA and the output of the $H(\cdot)$ be $n$ bits. Then, given a message $m$, the time complexity of finding $m'\neq m$ such that $H(m') = H(m)$ is $O(2^n)$ via brute-force search. This property implies that over a data exchange, if a malicious adversary aims to replace the actual message $m$ with a fake message $m'$ without being noticed ($H(m') = H(m)$ so that the receiver verifies the integrity of the message), and moreover if the adversary has a computational power of querying $2^\theta$ possible fake messages, then the probability of a successful fake message is $2^{\theta-n}$. Here we assume that the adversary knows the SHA so that it can check whether $H(m') = H(m)$ while trying different fake messages $m'$. The probability of success is negligible in practical settings where $n = 256$, $n = 512$, etc. For example, if an adversary can query $2^{80}$ fake messages and $n=256$, the probability of success is $2^{-176}$ for a single data package.

Furthermore, since the received digital signature can only be decrypted via the public key of the sender node, the receiver can verify the identity of the sender. Assuming (reasonably) that the private keys are kept secret and the digital signatures are $k$ bits, the time complexity of generating a successful fake signature is $O(2^k)$ via brute-force search \cite{nguyen2018survey}. Then, if an adversary does not know the public key of the sender (so that it cannot check whether a fake signature is decrypted via the public key), the probability of generating a successful fake signature for a chosen fake message is $2^{-k}$. On the other hand, obtaining public keys might be easier than private keys because the public keys are distributed over the network and all public keys can be accessed by hacking only one node. Then, if an adversary knows the public key of the sender, it can try different fake signatures for a chosen fake message and check whether the fake message is verified. In this case, if the adversary has the computational power of querying $2^\theta$ possible fake signatures, then the probability of success is $2^{\theta-k}$. Again choosing $k$ sufficiently high, such as $k = 256$, makes the success probability practically negligible. Notice that, however, if the private key of the sender is stolen, then the fake messages cannot be noticed at the receiver.

Over a data exchange, in case either the integrity of the received data package or the identity of the sender cannot be validated, then the received message is ignored and a retransmission can take place. Thereby, thanks to the asymmetric encryption procedure, the inter-node data exchanges are secured against attacks that manipulate either the message or the identity of the sender, such as man-in-the-middle and IP spoofing attacks. 

\subsubsection{Distributed Ledger and Consensus Mechanism} \label{sec:bc-consensus}
\begin{figure}[t]
\center
  \includegraphics[width=82mm]{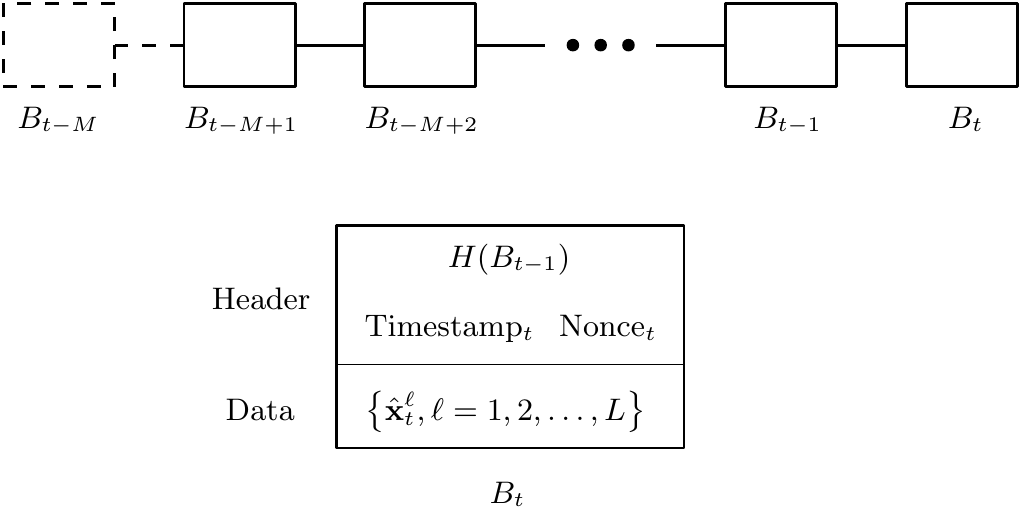}
\caption{\footnotesize The block structure and the distributed ledger at time $t$. $B_t$ denotes the block generated at time $t$ and $H(\cdot)$ denotes the cryptographic SHA. Each block contains a header and the estimates of the all local state vectors of the corresponding time. The ledger is composed of the $M$ recent blocks where the oldest block, $B_{t-M}$, is pruned at each time $t$.}
\label{fig:bc_structure}
\end{figure}
The ledger is a chronologically ordered sequence of blocks, stored at every node and synchronized over the network. In BC-based systems, the block content is application-specific. In our case (see Fig.~\ref{fig:bc_structure}), at each measurement sampling interval, a new block is produced, that includes (i) the state estimates of the current time and (ii) a header consisting of the discrete timestamp, hash value of the previous block that cryptographically links the current block to the previous block, and a random number called the nonce, which is the solution to a puzzle problem. Particularly, the nonce is determined, via brute-force search, such that the hash value of the current block satisfies a certain condition \cite{nguyen2018survey,GLiang18}. We next explain the process of producing a new block and the corresponding update in the distributed ledger.

At each time $t$, after computing the local state estimates, each node $\ell$ broadcasts a data package, containing its local state estimates, $\hat{\mathbf{x}}_t^{\ell}$, as the message, to the entire network. Then, as explained above, every node checks the validity of each received data package. A broadcasted data package is verified over the network only if the majority of the nodes validate it. After all the broadcasted data packages are verified, all the local state estimates of the current time are recorded into a new block at each node.

In this process, to alleviate inconsistencies in the distributed ledger, we need to make sure that the ledger is synchronously updated over the network. Moreover, we seek a mutual consensus over the network for the update of the ledger. Using the common Proof-of-Work consensus mechanism \cite{nguyen2018survey}, some nodes act as ``miners'', where the miners compete with each other to solve the puzzle and obtain the nonce value. Public BC-based systems such as Bitcoin \cite{Bitcoin} provides an incentive to miners, where the miner solving the puzzle first, gets a financial reward. In our case, however, this procedure is completely autonomous in that at each time, among all the local centers, a few nodes are randomly assigned as the miners. Then, the miner solving the puzzle first broadcasts the nonce value to the entire network. Each node then checks the puzzle solution. If the majority of nodes verify the solution, then the new block is produced and simultaneously connected to the existing ledger at every node. Here, random assignment of the miners enables a higher level of security to the BC mechanism compared to permanently assigning the miner nodes, that would then be the open targets for adversaries. 

As explained before, for a reliable dynamic state estimation, we require that the most recent state estimates are secure. Moreover, in case of an anomaly over the network, e.g., a failure or an attack, we can recover the system state from the recent reliable state estimates (the details are presented in Sec.~\ref{sec:robust} and Sec.~\ref{sec:trust}). For these reasons, in the distributed ledger, we propose to store a finite number of blocks that contain the recent state estimates in order to protect them against all kinds of manipulations. Let the number of blocks in the ledger be $M \geq 1$. Then, at each time, while a new block is connected to the existing ledger, the oldest block is pruned (see Fig.~\ref{fig:bc_structure}). This, in fact, solves the problem of monotonically increasing storage costs in the conventional BCs as well. We will explain how to choose $M$ in the subsequent sections. Further, in our case, at each time interval, only one new block is generated and we have a main chain of blocks without any forks, unlike the well-known public BCs such as Bitcoin \cite{Bitcoin} and Ethereum \cite{Ethereum}, in which the ledger keeps record of the transactions between nodes and it is possible to observe multiple transactions at the same time. Finally, we assume that the update of the ledger is completed within one measurement sampling interval.

The distributed ledger is resistant to modifications due to the following reasons: (i) since each block is cryptographically linked to the previous block, to modify a single block without being noticed, all the subsequent blocks must be modified accordingly, and (ii) since updating the ledger requires mutual consensus over the network, in order to modify the ledger, malicious entities need to control the majority of nodes in the network. This also implies that the security improvements introduced with the BC-based system design are valid only if the majority of the nodes in the network are reliable. We expect that this condition is easily satisfied in large-scale smart grids with many nodes distributed over a geographically wide region, for which hacking the majority of the nodes is practically quite difficult. From an attacker's perspective, letting $S \triangleq \lceil L/2 \rceil$ be the minimum number of nodes to be hacked in order to control the majority and hence to be able to arbitrarily modify the ledger, the best strategy would be attacking on the least costly set (in terms of the required efforts and resources for hacking) among the ${L}\choose{S}$ possible sets of nodes. Furthermore, an attacker may replace a block, say $B_t$, with a fake block $B_t'$ such that $H(B_t') = H(B_t)$ with a practically negligible probability (similar to the analysis in Sec.~\ref{sec:bc-data-exchange}), however, the fake block can still be noticed by comparing the modified ledger with the other copies of the ledger in the network. Hence, for a successful fake block, the attacker still needs to control the majority of the nodes.

{\textit{Remark 1:} Private BC mechanisms can be much more efficient compared to public BCs \cite{lombardi2018,GLiang18}. Moreover, recent research shows promising improvements in the speed of BC system operation, see e.g., \cite{super-fast-BC}, indicating that the BC efficiency can be safely improved in the near future, that enables using BC in time-sensitive real-time applications such as dynamic state estimation. Hence, in our mechanism, the update of the ledger can be completed within one measurement sampling interval using a fast private BC mechanism.}

\section{Distributed State Estimation Assuming Regular System Operation} \label{sec:SE-nominal}

In this section, assuming that the grid operation always fits to the nominal system model (see \eqref{eq:state_upd} and \eqref{eq:meas_model}), we aim to perform optimal state estimation in a fully-distributed manner, where each node $\ell$ only estimates its local state vector $\mathbf{x}_{t}^{\ell}$.

For an optimal state estimation, the node $\ell$ needs to use all the measurements that bear information about $\mathbf{x}_t^{\ell}$, either fully or partially. The node $\ell$ has access to only the local measurements, $\mathbf{y}_t^{\ell}$, that are clearly informative about $\mathbf{x}_t^{\ell}$ (see \eqref{eq:local_monitoring}). On the other hand, due to the shared state variables, some measurements collected at the other nodes may also be informative about $\mathbf{x}_t^{\ell}$.
Let
\begin{equation}\nonumber
\mathcal{C}^{\ell} \triangleq \{j \,|\, \mathbf{x}_t^{j} \cap \mathbf{x}_t^{\ell} \neq \emptyset, j \in \{1,\dots,L\} \backslash \{\ell\} \}
\end{equation}
be the set of nodes that share at least one state variable with the $\ell$th node.
Let $j \in \mathcal{C}^{\ell}$ and $\mathbf{y}_t^{\ell,j}$ be the sub-vector of $\mathbf{y}_t^{j}$ that is informative about $\mathbf{x}_t^{\ell}$, i.e.,
\begin{equation} \nonumber
\mathbf{y}_t^{\ell,j} \triangleq \{y_{k,t} \, | \, k \in \mathcal{R}^{j}, \mathcal{X}_{y_{k}} \cap \mathbf{x}_t^{\ell} \neq \emptyset\}.
\end{equation}
Moreover, let $\mathbf{x}_t^{\bar{\ell},j} \triangleq \mathbf{x}_t^{j} \backslash \mathbf{x}_t^{\ell}$. Then, we can decompose $\mathbf{y}_t^{\ell,j}$ as
\begin{gather} \label{eq:decomposition}
\mathbf{y}_t^{\ell,j} = \mathbf{H}^{\ell,j} \mathbf{x}_{t}^{\ell} + \mathbf{H}^{\bar{\ell},j} \mathbf{x}_t^{\bar{\ell},j} + \mathbf{w}_t^{\ell,j},
\end{gather}
where the matrices $\mathbf{H}^{\ell,j}$ and $\mathbf{H}^{\bar{\ell},j}$ are easily determined such that the equality in \eqref{eq:decomposition} is satisfied for all $t$. Moreover, $\mathbf{w}_t^{\ell,j}$ is the sub-vector of $\mathbf{w}_t$ corresponding to $\mathbf{y}_t^{\ell,j}$.

In \eqref{eq:decomposition}, the term $\mathbf{H}^{\bar{\ell},j} \mathbf{x}_t^{\bar{\ell},j}$ is clearly non-informative and irrelevant to the state estimator of the $\ell$th node. On the other hand, the $j$th node estimates $\mathbf{x}_{t}^{j}$ (and hence $\mathbf{x}_t^{\bar{\ell},j}$) at each time $t$. Then, denoting the estimate of $\mathbf{x}_t^{\bar{\ell},j}$ by $\mathbf{\hat{x}}_t^{\bar{\ell},j}$, based on \eqref{eq:decomposition}, we can write
\begin{align} \nonumber
{\tilde{\mathbf{y}}}_t^{\ell,j} &\triangleq \mathbf{y}_t^{\ell,j} - \mathbf{H}^{\bar{\ell},j} \mathbf{\hat{x}}_t^{\bar{\ell},j} \\ \nonumber
&= \mathbf{H}^{\ell,j} \mathbf{x}_{t}^{\ell} + \mathbf{H}^{\bar{\ell},j} (\mathbf{x}_t^{\bar{\ell},j} - \mathbf{\hat{x}}_t^{\bar{\ell},j}) + \mathbf{w}_t^{\ell,j} \\ \label{eq:processed_meas}
&= \mathbf{H}^{\ell,j} \mathbf{x}_{t}^{\ell} + {\tilde{\mathbf{w}}}_t^{\ell,j},
\end{align}
where
\begin{equation} \label{eq:w_tilde_ell_j}
{\tilde{\mathbf{w}}}_t^{\ell,j} \triangleq \mathbf{H}^{\bar{\ell},j} (\mathbf{x}_t^{\bar{\ell},j} - \mathbf{\hat{x}}_t^{\bar{\ell},j}) + \mathbf{w}_t^{\ell,j}.
\end{equation}
We propose that the $j$th node subtracts $\mathbf{H}^{\bar{\ell},j} \mathbf{\hat{x}}_t^{\bar{\ell},j}$ from $\mathbf{y}_t^{\ell,j}$ to compute ${\tilde{\mathbf{y}}}_t^{\ell,j}$ and then transmits it to the $\ell$th node at each time $t$ in order to facilitate the local state estimation at the $\ell$th node. Henceforth, we call ${\tilde{\mathbf{y}}}_t^{\ell,j}$ as the processed measurements (at the $j$th node for the $\ell$th node).

Notice that for each node $\ell$, \eqref{eq:local_states} defines the local state transition model. Moreover, the local measurements (see \eqref{eq:local_monitoring}) and the processed measurements to be received from the other nodes together form the overall measurement vector of the $\ell$th node (for the local state estimation task).
Let the overall measurement vector of the $\ell$th node be denoted with ${\tilde{\mathbf{y}}}_t^{\ell}$ and as an example, let $\mathcal{C}^{\ell} = \{i,j\}$. Then, ${\tilde{\mathbf{y}}}_t^{\ell}$ is simply obtained as follows:
\begin{equation}\nonumber
{\tilde{\mathbf{y}}}_t^{\ell} \triangleq
\begin{bmatrix}
\mathbf{y}_t^{\ell} \\
{\tilde{\mathbf{y}}}_t^{\ell,i} \\
{\tilde{\mathbf{y}}}_t^{\ell,j}
\end{bmatrix}.
\end{equation}
For the $\ell$th node, we then have the following linear state-space equations:
\begin{gather} \nonumber
\mathbf{x}_{t}^{\ell} = \mathbf{A}^{\ell} \mathbf{x}_{t-1}^{\ell} + \mathbf{v}_t^{\ell}, \\ \label{eq:local_measurements}
{\tilde{\mathbf{y}}}_t^{\ell} = {\tilde{\mathbf{H}}}^{\ell} \mathbf{x}_{t}^{\ell} + {\tilde{\mathbf{w}}}_t^{\ell},
\end{gather}
where ${\tilde{\mathbf{H}}}^{\ell}$ is determined based on $\mathbf{H}^{\ell}$, $\mathbf{H}^{\ell,i}$, and $\mathbf{H}^{\ell,j}$, and ${\tilde{\mathbf{w}}}_t^{\ell}$ is the noise vector corresponding to ${\tilde{\mathbf{y}}}_t^{\ell}$ (see \eqref{eq:local_monitoring} and \eqref{eq:processed_meas}):
\begin{equation}\label{eq:w_tilde}
{\tilde{\mathbf{H}}}^{\ell} \triangleq
\begin{bmatrix}
\mathbf{H}^{\ell} \\
\mathbf{H}^{\ell,i} \\
\mathbf{H}^{\ell,j}
\end{bmatrix}, ~
{\tilde{\mathbf{w}}}_t^{\ell} \triangleq
\begin{bmatrix}
\mathbf{w}_t^{\ell} \\
{\tilde{\mathbf{w}}}_t^{\ell,i} \\
{\tilde{\mathbf{w}}}_t^{\ell,j}
\end{bmatrix}.
\end{equation}

The Kalman filter is an iterative real-time estimator consisting of prediction and measurement update steps at each iteration. For the linear system given in \eqref{eq:local_measurements}, the following equations describe the Kalman filter iteration at time $t$, where $\hat{\mathbf{x}}_{t|t'}^{\ell}$ denotes the state estimates of the $\ell$th node at time $t$ ($t' = t-1$ for prediction and $t' = t$ for measurement update) {under the assumption that the correlation between state variables belonging to different local centers is zero, that leads to the independence of $\mathbf{x}_{t}^{\ell}$ from ${\tilde{\mathbf{w}}}_t^{\ell,i}$ and  ${\tilde{\mathbf{w}}}_t^{\ell,j}$ (and hence from ${\tilde{\mathbf{w}}}_t^{\ell}$):}

\textbf{Prediction}:
\begin{gather} \nonumber
\hat{\mathbf{x}}_{t|t-1}^{\ell} = \mathbf{A}^{\ell} \hat{\mathbf{x}}_{t-1|t-1}^{\ell}, \\ \label{eq:prediction}
\mathbf{P}_{t|t-1}^{\ell} = \mathbf{A}^{\ell} \mathbf{P}_{t-1|t-1}^{\ell} {\mathbf{A}}^{{\ell}^\mathrm{T}} + \sigma_v^2 \, \mathbf{I}_{N^{\ell}},
\end{gather}

\textbf{Measurement Update}:
\begin{gather} \nonumber
\mathbf{G}_{t}^{\ell} = \mathbf{P}_{t|t-1}^{\ell} {\tilde{\mathbf{H}}}^{{\ell}^\mathrm{T}} ({\tilde{\mathbf{H}}}^{\ell} \mathbf{P}_{t|t-1}^{\ell} {\tilde{\mathbf{H}}}^{{\ell}^\mathrm{T}} + \mathbf{R}_t^{\ell})^{-1}, \\ \nonumber
\hat{\mathbf{x}}_{t|t}^{\ell} = \hat{\mathbf{x}}_{t|t-1}^{\ell} + \mathbf{G}_{t}^{\ell} ({\tilde{\mathbf{y}}}_t^{\ell} - \tilde{\mathbf{H}}^{\ell} \hat{\mathbf{x}}_{t|t-1}^{\ell}), \\ \label{eq:meas_upd}
\mathbf{P}_{t|t}^{\ell} = \mathbf{P}_{t|t-1}^{\ell} - \mathbf{G}_{t}^{\ell} \tilde{\mathbf{H}}^{\ell} \mathbf{P}_{t|t-1}^{\ell},
\end{gather}
where $\mathbf{P}_{t|t-1}^{\ell}$ and $\mathbf{P}_{t|t}^{\ell}$ denote the estimates of the state covariance matrix of the $\ell$th node at time $t$ based on the measurements up to $t-1$ and $t$, respectively. Moreover, $\mathbf{G}_{t}^{\ell}$ is the Kalman gain matrix of the $\ell$th node at time $t$ and $\mathbf{R}_t^{\ell}$ denotes the covariance matrix of ${\tilde{\mathbf{w}}}_t^{\ell}$.

We now go back to the process of obtaining ${\tilde{\mathbf{y}}}_t^{\ell,j}$ at the $j$th node. We see through \eqref{eq:processed_meas} that this process contains estimation errors due to the term $\mathbf{x}_t^{\bar{\ell},j} - \mathbf{\hat{x}}_t^{\bar{\ell},j}$. Our aim is to statistically characterize the estimation errors in order to compute the statistics of ${\tilde{\mathbf{w}}}_t^{\ell,j}$ (see \eqref{eq:w_tilde_ell_j}), which is, in fact, required for an optimal state estimation at the $\ell$th node. Note that the $j$th node estimates its local state vector $\mathbf{x}_{t}^{j}$ (and hence $\mathbf{x}_t^{\bar{\ell},j}$) via its local Kalman filter. We propose that
\begin{equation} \nonumber
\mathbf{\hat{x}}_t^{\bar{\ell},j} = \hat{\mathbf{x}}_{t|t-1}^{\bar{\ell},j},
\end{equation}
where $\hat{\mathbf{x}}_{t|t-1}^{\bar{\ell},j}$ is the sub-vector of $\hat{\mathbf{x}}_{t|t-1}^{j}$ corresponding to $\mathbf{x}_t^{\bar{\ell},j}$. Then, the following lemma states the distribution of ${\tilde{\mathbf{w}}}_t^{\ell,j}$.

\textbf{Lemma 1:}
\begin{equation} \label{eq:dist_ell_j}
{\tilde{\mathbf{w}}}_t^{\ell,j} \sim \mathbf{\mathcal{N}}(\mathbf{0}, \pmb{\Delta}_t^{\ell,j}),
\end{equation}
where
\begin{equation}\label{eq:dist_ell_j_covariance}
\pmb{\Delta}_t^{\ell,j} \triangleq \mathbf{H}^{\bar{\ell},j} \mathbf{P}_{t|t-1}^{\bar{\ell},j} {\mathbf{H}}^{{\bar{\ell},j}^\mathrm{T}} + \sigma_w^2 \, \mathbf{I}_{K^{\ell,j}},
\end{equation}
where $\mathbf{P}_{t|t-1}^{\bar{\ell},j}$ is the estimate of the state covariance matrix of $\mathbf{\hat{x}}_t^{\bar{\ell},j}$, that can be obtained from $\mathbf{P}_{t|t-1}^{j}$ and $K^{\ell,j}$ denotes the size of the vector ${\tilde{\mathbf{w}}}_t^{\ell,j}$.

\begin{proof}
See Appendix~\ref{appendix:lemma1}.
\end{proof}

Through \eqref{eq:local_states} and \eqref{eq:local_monitoring}, we know that $\mathbf{v}_t^{\ell}$ and $\mathbf{w}_t^{\ell}$ are independent zero-mean multivariate Gaussian noise vectors. Furthermore, from Lemma 1, we know that the noise term ${\tilde{\mathbf{w}}}_t^{\ell,j}$ for the processed measurements (see \eqref{eq:processed_meas}) is also zero-mean multivariate Gaussian. That implies ${\tilde{\mathbf{w}}}_t^{\ell}$ is a zero-mean multivariate Gaussian vector (see \eqref{eq:w_tilde}). Then, since all the noise terms are Gaussian for the linear system in \eqref{eq:local_measurements}, the local Kalman filter given in \eqref{eq:prediction}--\eqref{eq:meas_upd} is the optimal state estimator for the $\ell$th node in minimizing the mean squared state estimation error \cite{Kalman_60}.

Based on Lemma 1, we can write (see \eqref{eq:w_tilde})
\begin{equation} \nonumber
{\tilde{\mathbf{w}}}_t^{\ell} \sim \mathbf{\mathcal{N}}(\mathbf{0},\mathbf{R}_t^{\ell}),
\end{equation}
where
\begin{align} \nonumber
\mathbf{R}_t^{\ell} &=
\begin{bmatrix}
  \mathrm{E}\big[\mathbf{w}_t^{\ell} \mathbf{w}_t^{{\ell}^\mathrm{T}}\big] & \mathrm{E}\big[\mathbf{w}_t^{\ell} {\tilde{\mathbf{w}}}_t^{{\ell,i}^\mathrm{T}}\big] & \mathrm{E}\big[\mathbf{w}_t^{\ell} {\tilde{\mathbf{w}}}_t^{{\ell,j}^\mathrm{T}}\big] \\
  \mathrm{E}\big[{\tilde{\mathbf{w}}}_t^{\ell,i} \mathbf{w}_t^{{\ell}^\mathrm{T}}\big] & \mathrm{E}\big[{\tilde{\mathbf{w}}}_t^{\ell,i} {\tilde{\mathbf{w}}}_t^{{\ell,i}^\mathrm{T}}\big] & \mathrm{E}\big[{\tilde{\mathbf{w}}}_t^{\ell,i} {\tilde{\mathbf{w}}}_t^{{\ell,j}^\mathrm{T}}\big] \\
  \mathrm{E}\big[{\tilde{\mathbf{w}}}_t^{\ell,j} \mathbf{w}_t^{{\ell}^\mathrm{T}}\big] & \mathrm{E}\big[{\tilde{\mathbf{w}}}_t^{\ell,j} {\tilde{\mathbf{w}}}_t^{{\ell,i}^\mathrm{T}}\big] & \mathrm{E}\big[{\tilde{\mathbf{w}}}_t^{\ell,j} {\tilde{\mathbf{w}}}_t^{{\ell,j}^\mathrm{T}}\big]
\end{bmatrix}
\\ \label{eq:compute_the_cov_matrix}
&= \begin{bmatrix}
  \sigma_w^2 \, \mathbf{I}_{K^{\ell}} & \mathbf{0} & \mathbf{0} \\
  \mathbf{0} & \pmb{\Delta}_t^{\ell,i} & \mathrm{E}\big[{\tilde{\mathbf{w}}}_t^{\ell,i} {\tilde{\mathbf{w}}}_t^{{\ell,j}^\mathrm{T}}\big] \\
  \mathbf{0} & \mathrm{E}\big[{\tilde{\mathbf{w}}}_t^{\ell,j} {\tilde{\mathbf{w}}}_t^{{\ell,i}^\mathrm{T}}\big] & \pmb{\Delta}_t^{\ell,j}
\end{bmatrix}.
\end{align}

Here, we observe that the computation of $\mathbf{R}_t^{\ell}$ requires the correlation between $\tilde{\mathbf{w}}_t^{\ell,i}$ and $\tilde{\mathbf{w}}_t^{\ell,j}$. Using \eqref{eq:w_tilde_ell_j}, we can write
\begin{align}\nonumber
\mathrm{E}\Big[{\tilde{\mathbf{w}}}_t^{\ell,i} {\tilde{\mathbf{w}}}_t^{{\ell,j}^\mathrm{T}}\Big] &= \mathbf{H}^{\bar{\ell},i} \, \mathrm{E}\Big[ \big(\mathbf{x}_t^{\bar{\ell},i} - \mathbf{\hat{x}}_t^{\bar{\ell},i}\big) \big(\mathbf{x}_t^{\bar{\ell},j} - \mathbf{\hat{x}}_t^{\bar{\ell},j}\big)^{\scriptscriptstyle \mathrm{T}} \Big] {\mathbf{H}}^{{\bar{\ell},j}^{\mathrm{T}}},
\end{align}
which requires the correlation between the state estimation errors for $\mathbf{x}_t^{\bar{\ell},i}$ and $\mathbf{x}_t^{\bar{\ell},j}$. However, since it is possible that $\mathbf{x}_t^{\bar{\ell},\bar{j},i} \triangleq \mathbf{x}_t^{\bar{\ell},i} \backslash \mathbf{x}_t^{\bar{\ell},j} \neq \emptyset$ or $\mathbf{x}_t^{\bar{\ell},\bar{i},j} \triangleq \mathbf{x}_t^{\bar{\ell},j} \backslash \mathbf{x}_t^{\bar{\ell},i} \neq \emptyset$ and the correlations for such state variables that belong to different local centers are not computed over the network, we approximate the correlation terms involving the state variables in $\mathbf{x}_t^{\bar{\ell},\bar{j},i}$ and $\mathbf{x}_t^{\bar{\ell},\bar{i},j}$ as zero. {Notice that the same assumption is made in the local Kalman filter equations above and in fact, it is the only assumption consistently made in this study, that looses the optimality of the proposed solution.} Through simulations, we observe that it only slightly increases the state estimation error compared to the optimal centralized Kalman filter. Hence, in the rest of the design and analysis (in Sec.~\ref{sec:robust} and Sec.~\ref{sec:trust}), we assume that the proposed distributed dynamic state estimator achieves near-optimal performance. Finally, we note that the main contributions of the proposed distributed dynamic state estimator are in the processing of the measurements that are acquired at the other local centers $j \in \mathcal{C}^{\ell}$ and only partially relevant to the local state vector $\mathbf{x}_t^{\ell}$.

\textit{Remark 2:} Each node $\ell$ needs the knowledge of $\mathbf{R}_t^{\ell}$ for the measurement update step of its local Kalman filter, where computing $\mathbf{R}_t^{\ell}$ requires $\pmb{\Delta}_t^{\ell,j}, \forall j \in \mathcal{C}^{\ell}$ (see \eqref{eq:compute_the_cov_matrix}). We observe through \eqref{eq:dist_ell_j_covariance} that $\pmb{\Delta}_t^{\ell,j}$ depends on $\mathbf{H}^{\bar{\ell},j}$ and $\mathbf{P}_{t|t-1}^{\bar{\ell},j}$. Here, $\mathbf{H}^{\bar{\ell},j}$ is determined based on the network topology, which is available at every node. On the other hand, $\mathbf{P}_{t|t-1}^{\bar{\ell},j}$ is extracted from the state covariance matrix of the $j$th node, $\mathbf{P}_{t|t-1}^{j}$, which is primarily computed at the $j$th node. Nevertheless, we see through \eqref{eq:prediction} and \eqref{eq:meas_upd} that the (iterative) computation of $\mathbf{P}_{t|t-1}^{j}$ does not depend on online {sensor} measurements and hence it can be computed offline at each node in the network. Moreover, in the proposed distributed trust management scheme (see Sec.~\ref{sec:trust}), each node needs to compute the state covariance matrices of all other nodes. Hence, the proposed state estimation mechanism does not introduce further computational complexity beyond the trust management scheme.

\section{Secure State Estimation against Measurement Anomalies} \label{sec:robust}

The state estimator proposed in the previous section is based on the assumption that the network operation always fits to the nominal system model. However, in practice, various kinds of anomaly might appear all over the network, e.g., measurement anomalies, due to cyber-attacks or network faults. In order to achieve secure state estimation against measurement anomalies, we propose to quickly and reliably detect them and then eliminate their effects as much as possible.

Considering that the attackers can be advanced, strategic, or adaptive to the system and detector dynamics, it is hard to model all attack types \cite{Necip19,Necip19a}. Moreover, considering the complex cyber-physical nature of the smart grid, it is also difficult to model all types of network faults. Hence, in this study, we do not focus on particular anomaly types and rather we assume the anomaly type is totally unknown. On the other hand, once we detect an anomaly, this prevents us to recover the useful part of the anomalous measurements (if any). Our anomaly mitigation strategy is then to reject/neglect the anomalous measurements in the state estimation process and predict the system state, until the system is recovered back to the regular operating conditions, based on (i) the previous reliable state estimates, securely recorded in the distributed ledger and (ii) the nominal system model. As the {sensors} are distributed over the network, each node analyzes only its local measurements. We next explain the proposed measurement anomaly detection scheme at the $\ell$th node and then the corresponding state recovery over the network.

\subsection{Real-time Detection of Local Measurement Anomalies} \label{sec:pers_meas_det}

During the regular system operation, it is possible to observe infrequent outliers and the Kalman filter is known to be effective to compensate (suppress) small errors due to such infrequent outliers \cite{Khosravi16}. Hence, we are particularly interested in long-term anomalies where there exist a temporal relation between anomalous measurements. Our aim is to detect such anomalies timely and reliably using the measurements that become available sequentially over time. In this problem, although we can statistically characterize the nominal measurements sufficiently accurately based on the nominal system model and the online state estimates under regular system operation, measurements can take various unknown statistical forms in case of an anomaly. Hence, we follow a solution strategy in that we derive and monitor (over time) a univariate summary statistic that is informative about a possible deviation of the online measurements from their nominal model, as detailed next.

At each time $t$, the $\ell$th node locally observes $\mathbf{y}_t^{\ell}$ (see \eqref{eq:local_monitoring}). Moreover, based on the local Kalman filter, we have
\begin{gather}\label{eq:pred_err_ell}
\mathbf{x}_{t}^{\ell} - \hat{\mathbf{x}}_{t|t-1}^{\ell} \sim \mathbf{\mathcal{N}}(\mathbf{0},\mathbf{P}_{t|t-1}^{\ell}).
\end{gather}
Using \eqref{eq:local_monitoring} and \eqref{eq:pred_err_ell}, we can write
\begin{align} \nonumber
\mathbf{y}_t^{\ell} - \mathbf{H}^{\ell} \hat{\mathbf{x}}_{t|t-1}^{\ell} &= \mathbf{H}^{\ell} (\mathbf{x}_{t}^{\ell} - \hat{\mathbf{x}}_{t|t-1}^{\ell}) + \mathbf{w}_t^{\ell} \\ \label{eq:dist_y_ell}
&\sim \mathbf{\mathcal{N}}(\mathbf{0},\pmb{\Sigma}_t^{\ell}),
\end{align}
where
\begin{gather}\nonumber
\pmb{\Sigma}_t^{\ell} \triangleq {\mathbf{H}^{\ell}} \mathbf{P}_{t|t-1}^{\ell} {\mathbf{H}}^{{\ell}^\mathrm{T}} + \sigma_w^2 \, \mathbf{I}_{K^{\ell}}.
\end{gather}
Then, based on \eqref{eq:dist_y_ell},
\begin{gather}\label{eq:chi_ell}
\chi_t^{\ell} \triangleq \big(\mathbf{y}_t^{\ell} - \mathbf{H}^{\ell} \hat{\mathbf{x}}_{t|t-1}^{\ell}\big)^{\scriptscriptstyle \mathrm{T}} \, {\pmb{\Sigma}_t^{\ell}}^{-1} \big(\mathbf{y}_t^{\ell} - \mathbf{H}^{\ell} \hat{\mathbf{x}}_{t|t-1}^{\ell}\big)
\end{gather}
is a chi-squared random variable with $K^{\ell}$ degrees of freedom. Notice that $\chi_t^{\ell}$ has a time-invariant (stationary) distribution under regular system operation.

Let $F_{K^{\ell}}(\cdot)$ be the cumulative distribution function (cdf) of a chi-squared random variable with $K^{\ell}$ degrees of freedom. If the p-value (right tail probability) corresponding to $\chi_t^{\ell}$ satisfies
\begin{align}\label{eq:tail_prob}
p_t^{\ell} &\triangleq 1 - F_{K^{\ell}}(\chi_t^{\ell}) < \alpha,
\end{align}
then the corresponding local measurement vector $\mathbf{y}_t^{\ell}$ is considered as an outlier for the significance level of $\alpha$.
In case of an anomaly, we expect that the chi-squared statistic $\chi_t^{\ell}$ takes higher values compared to its nominal values and hence, we expect to observe more frequent outliers. Then, we can model an anomaly as persistent outliers, as in \cite{Necip19b} and \cite{Yilmaz17}.

Based on \eqref{eq:tail_prob}, for an outlier $\mathbf{y}_t^{\ell}$, we have
\begin{gather} \label{eq:s}
s_t^{\ell} \triangleq \log\left(\frac{\alpha}{p_t^{\ell}}\right) > 0,
\end{gather}
and similarly, for a non-outlier $\mathbf{y}_t^{\ell}$, we have
$s_t^{\ell} \leq 0$.
Hence, we can consider $s_t^{\ell}$ as a (positive/negative) statistical evidence for anomaly at time $t$. Then, similar to the accumulation of the log-likelihood ratios in the well-known cumulative sum (CUSUM) test, we can accumulate $s_t^{\ell}$'s over time and declare a measurement anomaly only if there is a strong/reliable evidence, that results in the following CUSUM-like test \cite{Necip19b}:
\begin{align} \nonumber
\Gamma^{\ell} &= \inf\{t: g_t^{\ell} \geq h\}, \\ \label{eq:persistent-detection}
g_t^{\ell} &= \max\{0, g_{t-1}^{\ell} + s_t^{\ell}\},
\end{align}
where $g_0^{\ell} = 0$ and $\Gamma^{\ell}$ denotes the stopping time at which an anomaly is detected at the $\ell$th node.

Let $\tau^{\ell}$ be the unknown change-point at which a local measurement anomaly happens at the $\ell$th node and continues thereafter. The CUSUM test always keeps the maximum likelihood estimate of the change-point and update it as the measurements become available over time \cite[Sec.~2.2]{Basseville93}. Let $\hat{\tau}^{\ell}$ be the change-point estimate of the proposed test. Initializing $\hat{\tau}^{\ell} = 0$ at $t=0$, whenever the decision statistic $g_t^{\ell}$ reaches zero, we make the following update:
$\hat{\tau}^{\ell} \gets t$.
In other words, $\hat{\tau}^{\ell}$ is the latest time-instant at which the decision statistic reaches zero. The final change-point estimate is determined when an anomaly is declared at the stopping time $\Gamma^{\ell}$. Hence, we have
\begin{equation}\nonumber
\hat{\tau}^{\ell} \triangleq \max\{t: g_t^{\ell} = 0, t < \Gamma^{\ell}\}.
\end{equation}
The change-point estimate will be useful for state recovery (see Sec.~\ref{sec:recovery}).

For the CUSUM-like test in \eqref{eq:persistent-detection}, to achieve a lower false alarm rate (equivalently a larger average false alarm period), the significance level $\alpha$ is chosen smaller and/or the test threshold $h$ is chosen higher, that, on the other hand, leads to larger detection delays (see \eqref{eq:s} and \eqref{eq:persistent-detection}). Let $\mathrm{E}_\infty[\Gamma^{\ell}]$ be the average false alarm period, i.e., the average stopping time in the absence of anomalies ($\tau^{\ell} = \infty$). The following corollary (to Theorem 2 of \cite{Necip19b}) describes how to choose $\alpha$ and $h$ to obtain a desired lower bound $L$ on the average false alarm period.

\textbf{Corollary 1:} For a chosen $\alpha < 1/e$ and
\begin{gather}\label{eq:fap_lower_bnd}
h \geq \frac{\log(L)}{1 - {W(\alpha \log(\alpha))}/{\log(\alpha)}},
\end{gather}
we have
\begin{gather}\nonumber
\mathrm{E}_\infty[\Gamma^{\ell}] \geq L,
\end{gather}
where $W(\cdot)$ denotes the Lambert-W function\footnote{There exists a built-in MATLAB function \textbf{lambertw}.}.

\begin{proof}
See Appendix~\ref{appendix:corr1}.
\end{proof}

\subsection{State Recovery} \label{sec:recovery}

Once the proposed CUSUM-like detection scheme in \eqref{eq:persistent-detection} declares an anomaly, our purpose is to recover the current (and the future) state estimates. Since the local measurements observed after the (unknown) change-point, i.e., $\mathbf{y}_t^{\ell}, t > \tau^{\ell}$, are not reliable, we can recover the state estimates from the latest reliable state estimates computed at the change-point estimate $\hat{\tau}^{\ell}$.

The smart grid is a highly interconnected network, as in the proposed mechanism, the local state estimation is performed using the local measurements as well as the processed measurements received from some other nodes. Hence, if an anomaly happens at a node, the whole network is affected by the anomaly to some extent. Then, whenever a measurement anomaly is detected at a node, say the $\ell$th one at time $t$, the $\ell$th node immediately broadcasts $\hat{\tau}^{\ell}$ to the entire network. Then every node $j \in \{1,\dots,L\}$ makes the following state recovery:
\begin{gather} \label{eq:recovery}
\hat{\mathbf{x}}_{t|t}^{j} = {\mathbf{A}^{j}}^{(t-\hat{\tau}^{\ell})} \hat{\mathbf{x}}_{\hat{\tau}^{\ell}|\hat{\tau}^{\ell}}^{j},
\end{gather}
that essentially corresponds to the case where we replace all the measurements $\tilde{\mathbf{y}}_t^{j}$ during the anomaly interval $(\hat{\tau}^{\ell},t]$ with the corresponding pseudo measurements $\mathbf{\hat{y}}_t^{j} \triangleq \tilde{\mathbf{H}}^{j} \hat{\mathbf{x}}_{t|t-1}^{j}$, that makes the measurement innovation signal zero (see \eqref{eq:prediction} and \eqref{eq:meas_upd}).

The regular network operation requires the participation of every node in the state estimation process.
Hence, whenever a measurement anomaly is detected at the $\ell$th node, we propose to raise an alarm flag, calling for further investigation at the $\ell$th subregion and the neighboring subregions $j \in \mathcal{C}^{\ell}$ considering the possibility that the processed measurements received from the neighboring nodes may also lead to an anomaly in the local state estimation process. The investigation should be performed considering also the possibility of false alarms. After the investigation process and a possible recovery of the system, the predesigned regular network operation is restarted. Our main purpose here is to decrease the state estimation errors due to anomalies and hence to provide more reliable state estimates during the anomaly mitigation/system recovery period.

\textit{Remark 3:}  For the state recovery, the distributed ledger consisting of the $M$ recent blocks needs to include the state estimates computed at the change-point estimate $\hat{\tau}^{\ell}$, where $\hat{\tau}^{\ell}$ is not known ahead of time. However, since we expect quick detection, we do not expect to observe a $\hat{\tau}^{\ell}$ that is quite far away to the stopping time $\Gamma^{\ell}$. Hence, in practice $M$ can be chosen reasonably small. In the case where the ledger does not contain the state estimates of the time $\hat{\tau}^{\ell}$, we can recover the state estimates from the oldest state estimates available in the ledger considering that they are more reliable compared to the other alternatives.

\section{Distributed Trust Management} \label{sec:trust}

In BC-based distributed networks, malicious adversaries may obtain illegitimate access to the system, e.g., via stealing the digital identity of some nodes, malware propagation, etc., \cite{YGuo16,bertino2017botnets}, and additionally some nodes may get faulty during the system operation. Moreover, as the network is fully-distributed, there is no centralized trusted node to check whether all nodes are safe and trustable, i.e., whether the nodes are functioning according to the predesigned network rules. Therefore, against the possibility of misbehaving nodes, we need a distributed trust management mechanism over the network, in which all nodes collectively verify the trustability of each node. Recall that every node knows (i) the nominal system model and the network configuration, and (ii) a finite history of recent state estimates of all the nodes stored in the shared distributed ledger (see Sec.~\ref{sec:bc-consensus}). Using only (i) and (ii), each node votes on the trustability of all other nodes. Then, at each time, trustability of each node is decided via majority-voting. We explain below how the $\ell$th node is evaluated by the other regular (non-misbehaving) nodes.

Suppose that at an unknown time $\eta^{\ell}$, an unexpected event happens at the $\ell$th node: the node gets faulty or an attacker hacks and takes control of the node. Then, we can no longer expect that the behavior of the $\ell$th node fits to its pre-defined regular operation. Furthermore, similar to the measurement anomalies, it is quite difficult to model the (anomalous) behavior of the $\ell$th node after time $\eta^{\ell}$. We would like to detect misbehaving nodes as quickly as possible to timely mitigate the corresponding effects on the state estimation process. Hence, for the evaluation of the $\ell$th node, we propose that each node decides whether the state estimates provided by the $\ell$th node exhibit an anomalous pattern over time. In this direction, we next derive the nominal evolution (over time) model of the local state estimates of the $\ell$th node. Then, similar to Sec.~\ref{sec:pers_meas_det}, we derive a univariate summary statistic that is informative about a possible deviation of the local state estimates of the $\ell$th node from their nominal evolution model and monitor this statistic over time.

Based on the local Kalman filter iteration of the $\ell$th node at time $t$ (see \eqref{eq:prediction} and \eqref{eq:meas_upd}), we can write
\begin{align}\nonumber
\hat{\mathbf{x}}_{t|t}^{\ell} &= \hat{\mathbf{x}}_{t|t-1}^{\ell} + \mathbf{G}_{t}^{\ell} \big({\tilde{\mathbf{y}}}_t^{\ell} - {\tilde{\mathbf{H}}}^{\ell} \hat{\mathbf{x}}_{t|t-1}^{\ell}\big) \\ \label{eq:tmp2}
&= \mathbf{A}^{\ell} \hat{\mathbf{x}}_{t-1|t-1}^{\ell} + \mathbf{G}_{t}^{\ell}\big({\tilde{\mathbf{H}}}^{\ell} (\mathbf{x}_{t}^{\ell} -  \hat{\mathbf{x}}_{t|t-1}^{\ell}) + {\tilde{\mathbf{w}}}_t^{\ell}\big) \\ \label{eq:evolution-state-estimate}
&\sim \mathbf{\mathcal{N}}\big(\mathbf{A}^{\ell} \hat{\mathbf{x}}_{t-1|t-1}^{\ell},\pmb{\Psi}_t^{\ell} \, \big),
\end{align}
where
\begin{gather}\label{eq:Gamma_t_ell}
\pmb{\Psi}_t^{\ell} \triangleq \mathbf{G}_{t}^{\ell} \big({\tilde{\mathbf{H}}}^{\ell} \mathbf{P}_{t|t-1}^{\ell}  {\tilde{\mathbf{H}}}^{{\ell}^\mathrm{T}} + \mathbf{R}_t^{\ell}\big) {\mathbf{G}}_{t}^{{\ell}^\mathrm{T}}.
\end{gather}
Here, \eqref{eq:tmp2} is obtained using \eqref{eq:local_measurements} and \eqref{eq:prediction}. Moreover, $\pmb{\Psi}_t^{\ell}$ is obtained using $\mathbf{x}_{t}^{\ell} - \hat{\mathbf{x}}_{t|t-1}^{\ell} \sim \mathbf{\mathcal{N}}(\mathbf{0},\mathbf{P}_{t|t-1}^{\ell})$ (see \eqref{eq:pred_err_ell}), ${\tilde{\mathbf{w}}}_t^{\ell} \sim \mathbf{\mathcal{N}}(\mathbf{0},\mathbf{R}_t^{\ell})$, and approximating the correlation terms involving the state variables belonging to different local centers as zero, as in Sec.~\ref{sec:SE-nominal}.

Notice that \eqref{eq:evolution-state-estimate} statistically characterizes the local state estimates at time $t$, given the local state estimates at time $t-1$, under regular system operation. Based on \eqref{eq:evolution-state-estimate}, we can write
\begin{equation}\nonumber
\hat{\mathbf{x}}_{t|t}^{\ell} - \mathbf{A}^{\ell} \hat{\mathbf{x}}_{t-1|t-1}^{\ell} \sim \mathbf{\mathcal{N}}(\mathbf{0},\pmb{\Psi}_t^{\ell}),
\end{equation}
that implies
\begin{equation}\label{eq:chi-sq-evolution}
\pi_t^{\ell} \triangleq \big(\hat{\mathbf{x}}_{t|t}^{\ell} - \mathbf{A}^{\ell} \hat{\mathbf{x}}_{t-1|t-1}^{\ell}\big)^{\scriptscriptstyle \mathrm{T}} \, {\pmb{\Psi}_t^{\ell}}^{-1} \big(\hat{\mathbf{x}}_{t|t}^{\ell} - \mathbf{A}^{\ell} \hat{\mathbf{x}}_{t-1|t-1}^{\ell}\big)
\end{equation}
is a chi-squared random variable with $N^{\ell}$ degrees of freedom under regular system operation.

If node $\ell$ is misbehaving, we expect that the local state estimates provided by it deviate from the nominal evolution model in \eqref{eq:evolution-state-estimate}, that makes $\pi_t^{\ell}$ larger than its nominal values. Then, similar to the real-time detection of measurement anomalies in Sec.~\ref{sec:pers_meas_det}, the following CUSUM-like test can be employed at a regular node $j$ to decide on the trustability of the $\ell$th node:
\begin{align} \nonumber
\Gamma_j^{\ell} &= \inf\{t: g_{j,t}^{\ell} \geq h^{\ell}\}, ~ g_{j,t}^{\ell} = \max\{0, g_{j,t-1}^{\ell} + s_{j,t}^{\ell}\}, \\ \label{eq:trust-detection}
s_{j,t}^{\ell} &= \log\left({\alpha^{\ell}}/{p_{j,t}^{\ell}}\right), ~ p_{j,t}^{\ell} = 1 - F_{N^{\ell}}(\pi_t^{\ell}),
\end{align}
where $g_{j,t}^{\ell}$ is the decision statistic at time $t$, $g_{j,0}^{\ell} = 0$, $h^{\ell}>0$ is the test threshold, $s_{j,t}^{\ell}$ is the statistical evidence (for misbehavior) at time $t$, $\alpha^{\ell}$ is the significance level, $p_{j,t}^{\ell}$ is the p-value corresponding to $\pi_t^{\ell}$, and $F_{N^{\ell}}(\cdot)$ denotes the cdf of a chi-squared random variable with $N^{\ell}$ degrees of freedom. A regular node $j$ then evaluates the $\ell$th node as trusted until time $\Gamma_j^{\ell}$, and misbehaving after $\Gamma_j^{\ell}$. Furthermore, as before, the unknown change-point $\eta^{\ell}$ is estimated by the $j$th node as the latest time-instant at which the decision statistic reaches zero before the stopping time $\Gamma_j^{\ell}$:
\begin{equation}\nonumber
\hat{\eta}_j^{\ell} \triangleq \max\{t: g_{j,t}^{\ell} = 0, t < \Gamma_j^{\ell}\}.
\end{equation}

The overall decision on the trustability/misbehavior of the $\ell$th node is made by the majority of the nodes. Let the vote of the $j$th node on the trustability of the $\ell$th node at time $t$ be denoted with a binary variable $d_{j,t}^{\ell}$, where $d_{j,t}^{\ell} = 0$ or $d_{j,t}^{\ell} = 1$ if the $j$th node evaluates the $\ell$th node as trusted or misbehaving, respectively. Notice that if the $j$th node is regular, then it votes based on the designed test in \eqref{eq:trust-detection} that gives rise to
\begin{equation}\nonumber
d_{j,t}^{\ell} = \mathbb{I}(\Gamma_j^{\ell} \leq t).
\end{equation}
The time at which the $\ell$th node is declared misbehaving over the network is then determined as follows:
\begin{equation}\label{eq:consensus}
\Gamma^{\ell}_{\text{net}} \triangleq \inf\bigg\{t: \sum_{j \, \in \, \{1,\dots,L\} \backslash \{\ell\}} d_{j,t}^{\ell} > \frac{L-1}{2}\bigg\}.
\end{equation}
Notice that this decision mechanism works as intended unless the majority of the nodes are misbehaving. In other words, as long as the majority of the nodes regularly employ the proposed misbehavior detection scheme in \eqref{eq:trust-detection} and vote accordingly, then the trustability of the $\ell$th node is reliably evaluated over the network, for every node $\ell$.

Under the nominal system operation, all nodes are regular and hence the misbehavior detector (for the $\ell$th node) in \eqref{eq:trust-detection} is identical at all the nodes $j$ in the network. Then, the false alarm rate of \eqref{eq:consensus} is equal to the false alarm rate of \eqref{eq:trust-detection}. Hence, the proposed trust management scheme achieves the same false alarm guarantees through Corollary 1 (after replacing the parameters $\alpha$ and $h$ in the corollary with $\alpha^{\ell}$ and $h^{\ell}$, respectively).

If \eqref{eq:consensus} declares the $\ell$th node as misbehaving, an alarm flag is raised, calling for an investigation at the $\ell$th node and the neighboring nodes $j \in \mathcal{C}^{\ell}$. Moreover, due to the inter-node data exchanges, a node misbehavior affects all the nodes in the network to some extent. Then, as before, until the system is recovered back to the regular operating conditions, the local states of each node $j \in \{1,\dots,L\}$ can be predicted based on the nominal system model and the latest reliable estimates computed at the change-point estimate $\hat{\eta}_j^{\ell}$ as follows:
\begin{gather}\nonumber
\hat{\mathbf{x}}_{t|t}^{j} = {\mathbf{A}^{j}}^{(t-\hat{\eta}_j^{\ell})} \hat{\mathbf{x}}_{\hat{\eta}_j^{\ell}|\hat{\eta}_j^{\ell}}^{j},
\end{gather}
where $\hat{\mathbf{x}}_{\hat{\eta}_j^{\ell}|\hat{\eta}_j^{\ell}}^{j}$ can be obtained from the distributed ledger.

The proposed trust management scheme requires that each node $j \in \{1,\dots,L\} \backslash \{\ell\}$ computes $\pi_t^{\ell}$ and $\pmb{\Psi}_t^{\ell}$ for every other node $\ell \in \{1,\dots,L\} \backslash \{j\}$ (see \eqref{eq:Gamma_t_ell}--\eqref{eq:consensus}). Since the nominal system model is known by every node and the local state estimates provided by the other nodes are available (via the distributed ledger) to every node, the node $j$ already knows
$\mathbf{A}^{\ell}$, ${\tilde{\mathbf{H}}}^{\ell}$, $\hat{\mathbf{x}}_{t|t}^{\ell}$, and $\hat{\mathbf{x}}_{t-1|t-1}^{\ell}$ at each time $t$ for every node $\ell$. On the other hand, $\mathbf{R}_t^{\ell}$, $\mathbf{G}_t^{\ell}$, and $\mathbf{P}_{t|t-1}^{\ell}$ are not directly available to the $j$th node. Fortunately, the Kalman gain matrix $\mathbf{G}_t^{\ell}$ and the estimate of the state covariance matrix $\mathbf{P}_{t|t-1}^{\ell}$ can be computed offline without requiring online {sensor} measurements. Moreover, $\mathbf{R}_t^{\ell}$ is computed based on the estimates of the state covariance matrices (see also Remark 2). Hence, at each node $j$, we propose to compute $\mathbf{G}_t^{\ell}$ and $\mathbf{P}_{t|t-1}^{\ell}$ (iteratively through \eqref{eq:prediction} and \eqref{eq:meas_upd}) for every other node $\ell$ in the network.

{\textit{Remark 4:} Both the local measurement anomaly detection and the distributed trust management schemes depend on a CUSUM-like anomaly detection algorithm, whose general form is first presented in \cite{Necip19b} along with an asymptotic performance analysis. On the other hand, the summary statistics used in the both schemes are designed in a novel task-specific manner. Further, although voting-based consensus is not new, achieving consensus based on the real-time misbehavior detection algorithms employed at every regular node is novel. Finally, the proposed state recovery scheme is novel, as after an anomaly/misbehavior is detected in real-time based on the sequence of the derived summary statistics, the change-point is estimated and then the state estimates are recovered accordingly from the latest reliable state estimates without assuming any anomaly/misbehavior model.}

\section{Summary of the Proposed Mechanism} \label{sec:summary}

\begin{figure}[t]
\center
  \includegraphics[width=90mm]{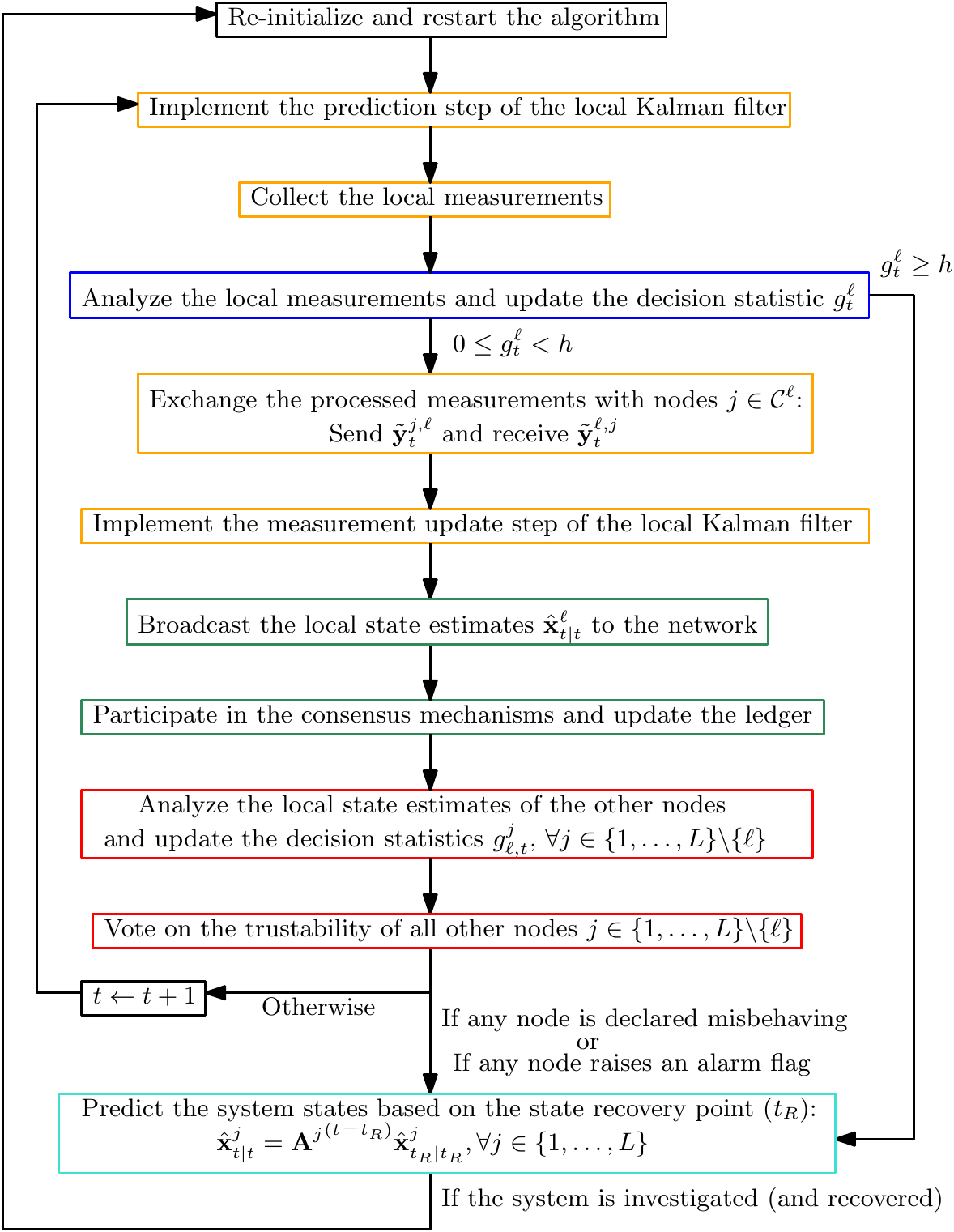}
\caption{\footnotesize Flowchart of the proposed procedure at the $\ell$th node. The green, orange, blue, and red rectangles mainly refer to the components of the proposed solution design discussed in Sec.~\ref{sec:blockchain}, Sec.~\ref{sec:SE-nominal}, Sec.~\ref{sec:robust}, and Sec.~\ref{sec:trust}, respectively. Moreover, the turquoise rectangle refers to the state recovery mechanism discussed in Sec.~\ref{sec:robust} and Sec.~\ref{sec:trust}.}
 \label{fig:flowchart}
\end{figure}

We summarize the proposed procedure at the $\ell$th node in Fig.~\ref{fig:flowchart}, where the procedure is identical at every node. Since the proposed mechanism requires an investigation after the detection of a measurement anomaly at any node or a misbehavior of any node, the overall stopping time of the network is given by
\begin{equation}\label{eq:net_stop_time}
\Gamma_{\text{net}} \triangleq \inf\{\Gamma^{\ell},\Gamma^{\ell}_{\text{net}},\ell = 1,\dots,L\}.
\end{equation}
If several detection mechanisms simultaneously give alarms, we can recover the state estimates based on the oldest among the corresponding change-point estimates. Moreover, if the state estimates corresponding to the (oldest) change-point estimate are not included in the distributed ledger consisting of the $M$ recent blocks, then we can choose $t_R = t-M+1$ as the state recovery point, that corresponds to the oldest state estimates available in the ledger, where the state recovery point is denoted with $t_R$ in Fig.~\ref{fig:flowchart}. In case an anomaly is declared over the network, after an investigation and possibly the recovery of the system, the proposed mechanism is restarted. 

\section{Simulation Results} \label{sec:numerical}

\begin{figure}[t]
\center
  \includegraphics[width=67mm]{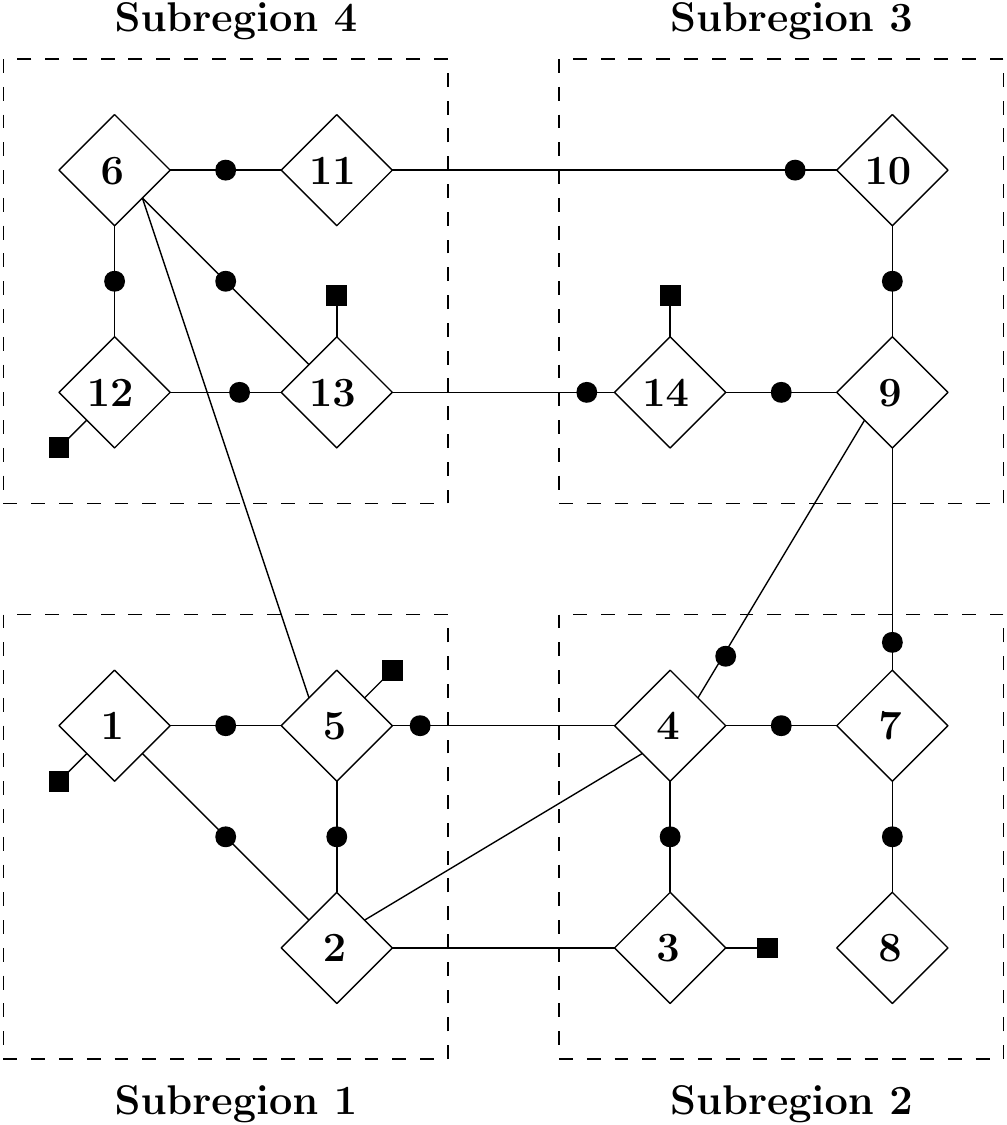}
\caption{\footnotesize Wide-area monitoring model of the IEEE-14 bus power system. Four subregions are illustrated with dashed rectangles. The circles on the branches represent the power flow measurements, and the squares represent the power injection measurements.}
 \label{fig:ieee14bus}
\end{figure}

In this section, we evaluate the performance of the proposed mechanism via simple case studies over an IEEE-14 bus power system. {We consider a special case where system state includes the voltage phase angles only and the measurements are the real power flow and power injection
measurements (assuming the voltage magnitudes are known).} The system consists of $4$ subregions, $N = 14$ buses, and $K = 23$ {sensors} (see Fig.~\ref{fig:ieee14bus}). The bus $6$ is chosen as the reference bus, the state transition matrix is chosen to be an identity matrix, and the measurement matrix $\mathbf{H}$ is determined based on the network topology. The noise variances are chosen as $\sigma_v^2 = \sigma_w^2 = 10^{-4}$, and the initial state variables (voltage phase angles) are determined via the DC optimal power flow algorithm for case-14 in MATPOWER \cite{Zimmerman11}. For the proposed detection schemes, to achieve $\mathrm{E}_\infty[\Gamma^{\ell}],\mathrm{E}_\infty[\Gamma^{\ell}_{\text{net}}] \geq 10^6, \forall \ell \in \{1,2,3,4\}$, we choose $\alpha = \alpha^1 = \dots = \alpha^4 = 0.2$ and $h = h^1 = \dots = h^4 = 21.3527$ (see \eqref{eq:fap_lower_bnd}). Then, we obtain via a Monte Carlo simulation that the average false alarm period of the network (see \eqref{eq:net_stop_time}) is $\mathrm{E}_\infty[\Gamma_{\text{net}}] \backsimeq 1.26\times10^6$. Moreover, we choose the number of blocks in the distributed ledger as $M = 200$. In the following, we present simulation results firstly for a measurement anomaly case and then for a node misbehavior case.

\subsection{Case 1: Measurement Anomaly}

As an example, we consider FDI attacks against the sensor measurements in the subregion $\ell$:
\begin{equation}\label{eq:FDI-attack}
\mathbf{y}_t^{\ell} = \mathbf{H}^{\ell} \mathbf{x}_{t}^{\ell} + \mathbf{a}_t^{\ell} + \mathbf{w}_t^{\ell}, ~ t \geq \tau^{\ell},
\end{equation}
where $\mathbf{a}_t^{\ell} = [a_{1,t}^{\ell}, \dots, a_{K{^\ell},t}^{\ell}]^{\scriptscriptstyle \mathrm{T}}$ denotes the injected false data at time $t$. We assume that subregions 1 and 2 are under FDI attack after time $\tau \triangleq \tau^{1} = \tau^{2}$ with $a_{i,t}^{\ell} \sim \mathcal{U}[0,\rho]$, $\forall i \in \{1,\dots,K^{\ell}\}$, $\forall \ell \in \{1,2\}$, $\forall t \geq \tau$, where $\mathcal{U}[0,\rho]$ denotes a uniform random variable in the range of $[0,\rho]$.

\begin{figure}[t]
\center
  \includegraphics[width=88mm]{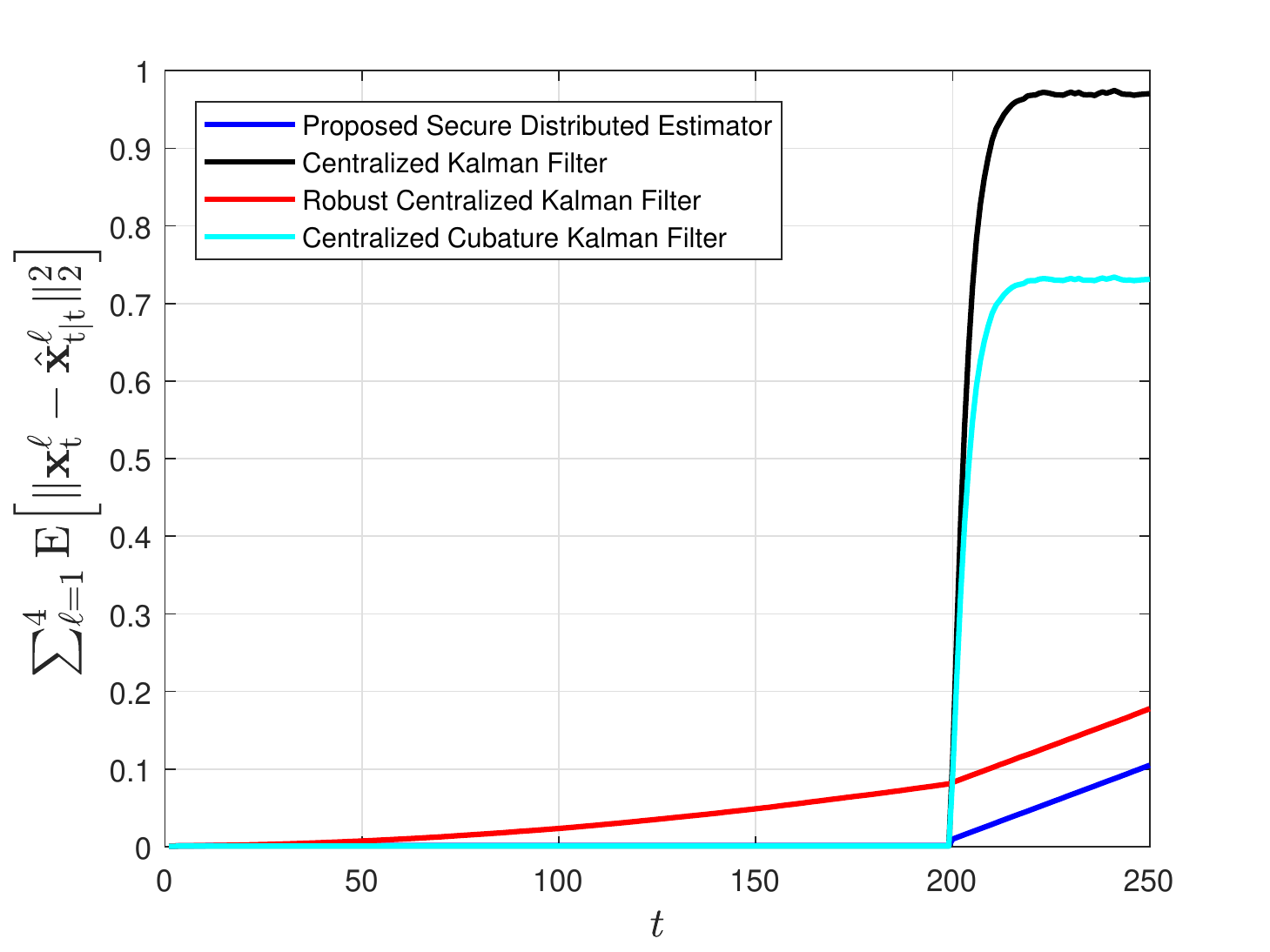}
\caption{\footnotesize Mean squared state estimation error vs. time in case of an FDI attack (with $\rho=0.3$) performed after time $\tau = 200$ targeting the {sensor} measurements in the subregions 1 and 2.}
 \label{fig:MSE_vs_time_meas_anom}
\end{figure}

Firstly, assuming $\tau = 200$ and $\rho = 0.3$, we present in Fig.~\ref{fig:MSE_vs_time_meas_anom} the sum of the mean squared state estimation errors over all local centers for both the pre-attack period, i.e., $t < 200$, and for the attacking period of $t \in [200, 250]$. We present the performance of the proposed distributed secure state estimation mechanism, the centralized Kalman filter, {the centralized cubature Kalman filter \cite{Haykin09} (since it is reported to be robust against attacks \cite{Taha18} although mainly designed for nonlinear systems),} and the robust centralized Kalman filter that rejects gross outliers and replaces the corresponding measurements with the pseudo measurements, similar to \cite{da1987efficient}. In particular, the robust Kalman filter computes a chi-squared statistic at each time $t$ using all the measurements $\mathbf{y}_t$, similar to \eqref{eq:chi_ell}, and compute the corresponding p-value based on the chi-squared distribution with $K$ degrees of freedom. The significance level for outliers is chosen as $0.01$. Then, if the p-value is less than $0.01$, the corresponding measurements $\mathbf{y}_t$ are replaced with the pseudo measurements $\hat{\mathbf{y}}_t \triangleq \mathbf{H} \hat{\mathbf{x}}_{t|t-1}$ in the robust centralized Kalman filter where $\hat{\mathbf{x}}_{t|t-1}$ denotes the predicted states at time $t$. We observe via Fig.~\ref{fig:MSE_vs_time_meas_anom} that the proposed estimator performs near-optimally in the pre-attack period and reduces the state estimation errors in the attacking period. Since the proposed mechanism does not use any measurements after the anomaly detection, state estimation/prediction error increases over time in the attacking period, as expected. Hence, after anomaly detection, the anomaly needs to be quickly mitigated to prevent large errors. {We also observe that although the cubature Kalman filter shows some performance improvement over the Kalman filter in the attacking regime, it still significantly deviates from the actual system state after the attack is being launched. Moreover, although the robust Kalman filter reduces errors over the attacking period, it deviates from the system state under regular conditions due to the rejection of nominal outliers.}

\begin{figure}[t]
\center
  \includegraphics[width=88mm]{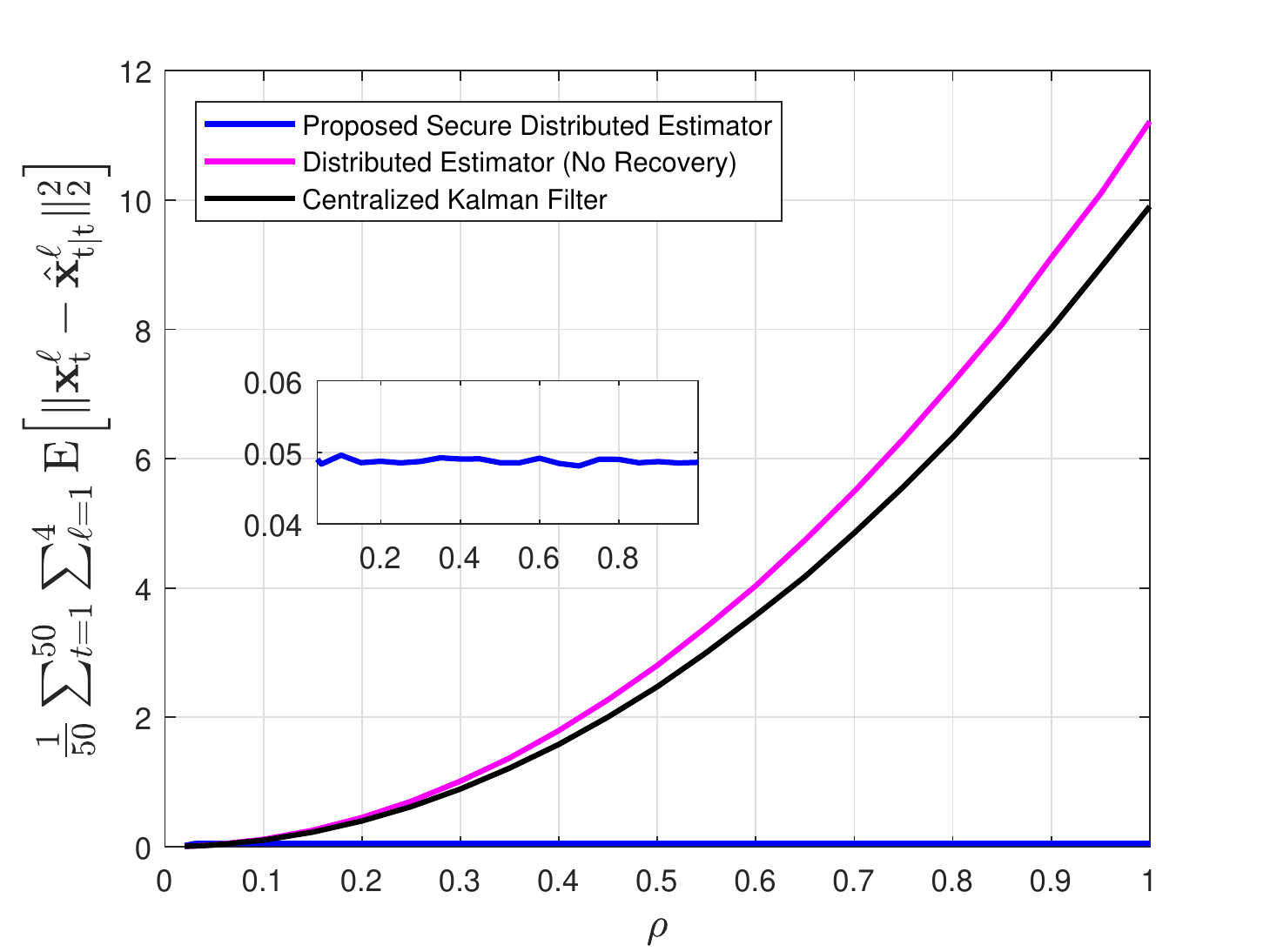}
\caption{\footnotesize Average state estimation error over an attacking period vs. attack magnitudes in case of FDI attacks to the {sensor} measurements in the subregions 1 and 2.}
 \label{fig:MSE_vs_magnitudes_meas}
\end{figure}

Next, we present in Fig.~\ref{fig:MSE_vs_magnitudes_meas} the average state estimation performance over an attacking period as the FDI attack magnitude $\rho$ varies. Particularly, we assume $\tau = 1$ and compute the sum of the mean squared state estimation errors over local centers, averaged over an attacking period of $t \in [1,50]$. Recall that after the detection of measurement anomalies and until the system is back to the regular conditions, the proposed mechanism predicts the system state based on the nominal system model and the previous reliable state estimates. Hence, after the detection, the state estimation/prediction performance of the proposed mechanism does not depend on the attack magnitudes. On the other hand, for the centralized Kalman filter and the distributed state estimator designed assuming ideal operating conditions (see Sec.~\ref{sec:SE-nominal}), the state estimation error increases with the attack magnitude as there are no anomaly detection/state recovery mechanisms integrated into them.

\begin{figure}[t]
\center
  \includegraphics[width=88mm]{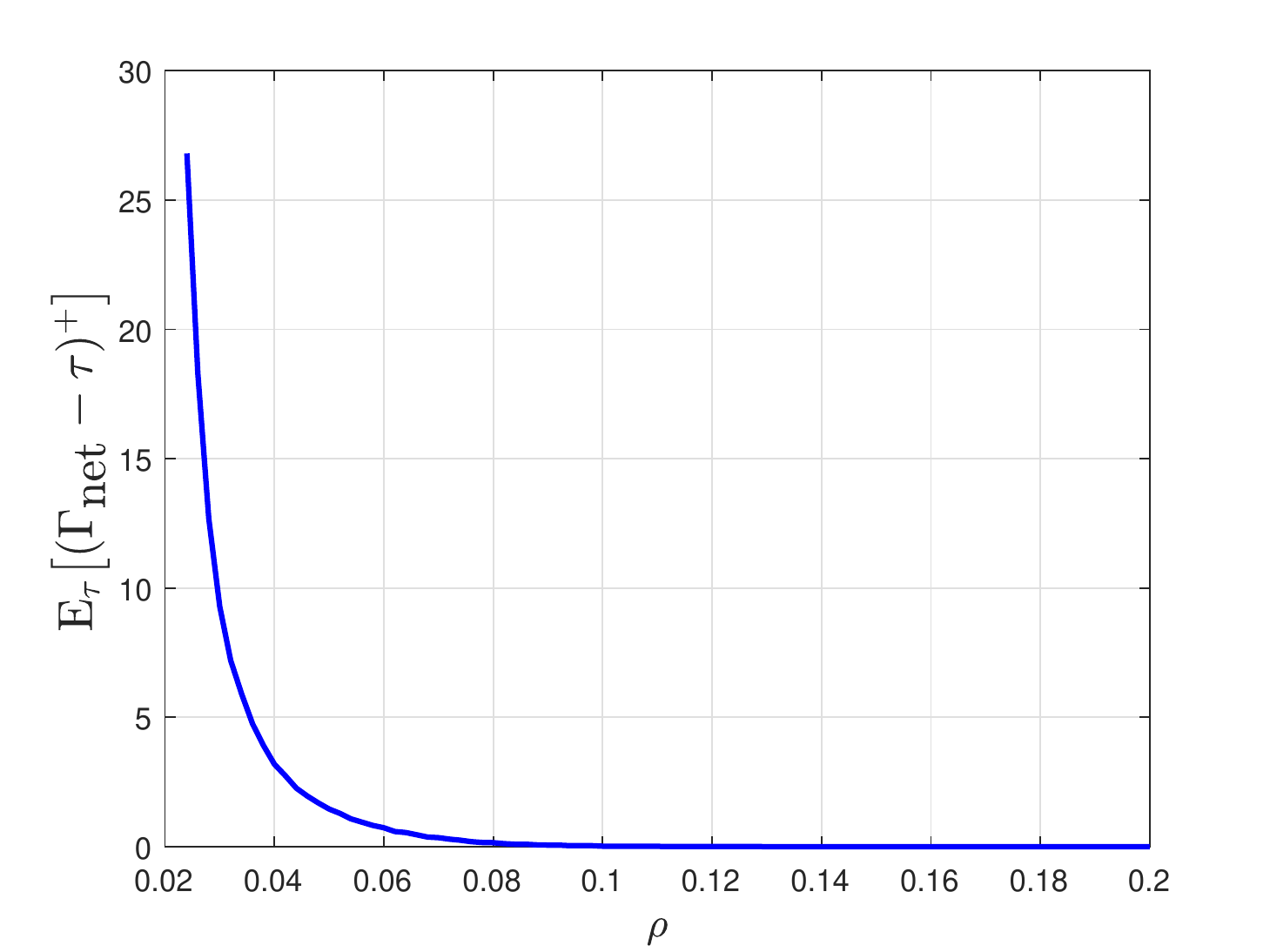}
\caption{\footnotesize Average detection delay vs. attack magnitudes in case of FDI attacks targeting the subregions 1 and 2.}
 \label{fig:delay_vs_mag_meas}
\end{figure}

Finally, we illustrate in Fig.~\ref{fig:delay_vs_mag_meas} how the average detection delay varies as the attack magnitude/intensity changes, assuming $\tau = 1$, that results in the worst-case detection delays since $g_t = 0$ at $t = 0$. As expected, for stronger attacks, we observe more significant outliers (smaller p-values) that leads to quicker detection (see \eqref{eq:s} and \eqref{eq:persistent-detection}).

\subsection{Case 2: Node Misbehavior}

\begin{figure}[t]
\center
  \includegraphics[width=88mm]{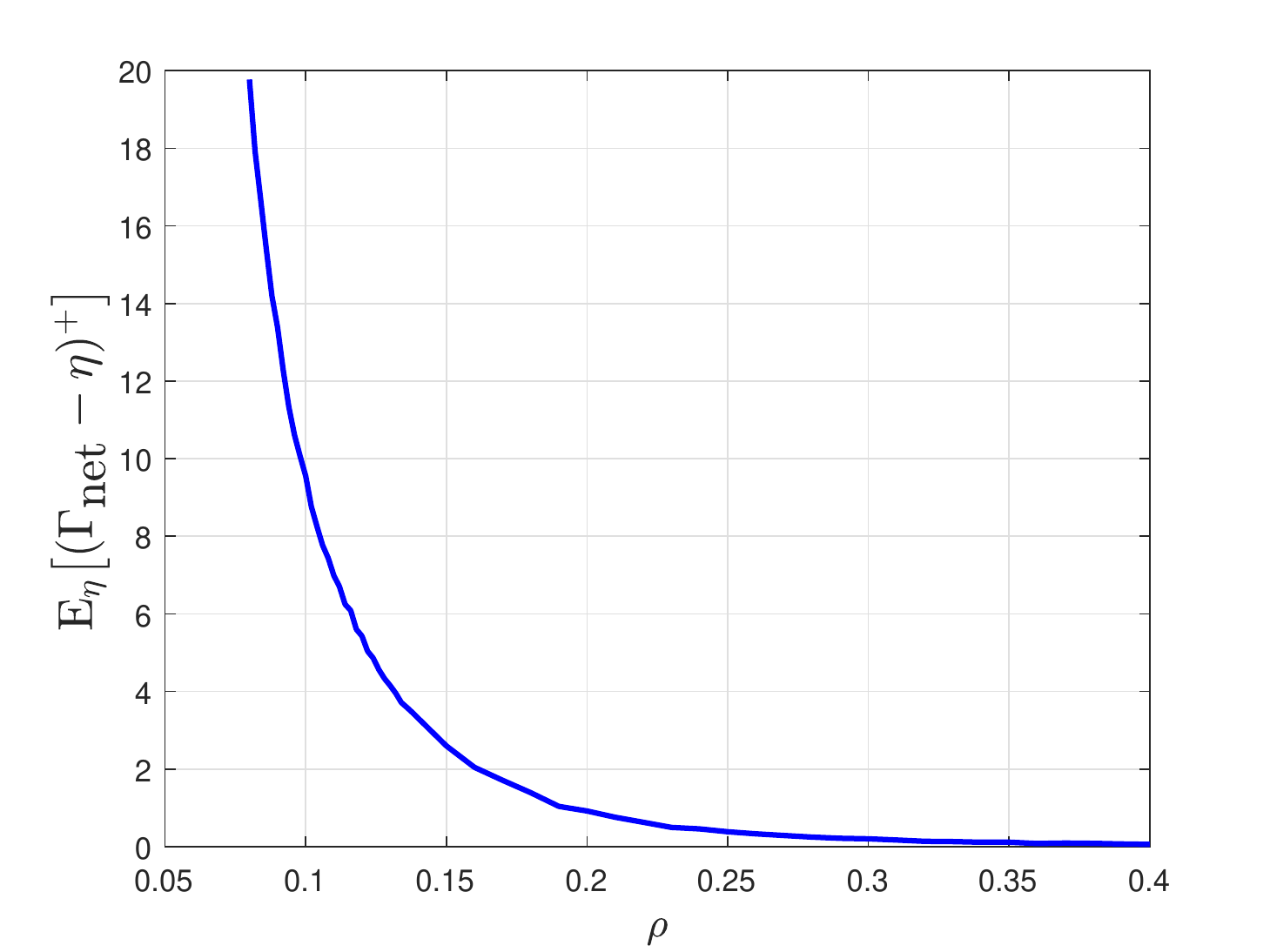}
\caption{\footnotesize Average detection delay vs. attack magnitudes. False data are injected at the local measurements of the node 3 and no anomaly detection mechanism is employed at the node 3.}
 \label{fig:delay_vs_mag_misb}
\end{figure}

\begin{figure}[t]
\center
  \includegraphics[width=88mm]{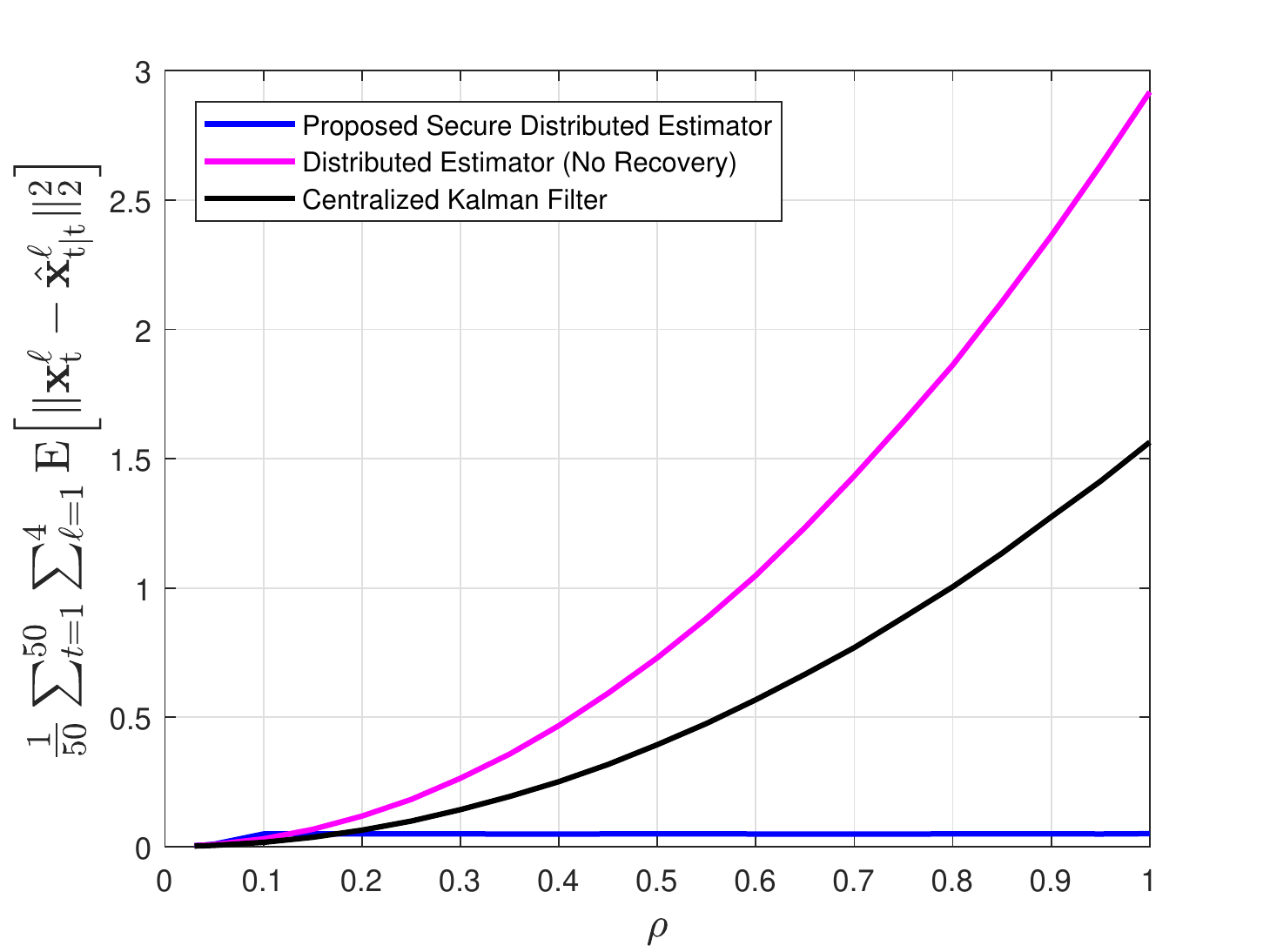}
\caption{\footnotesize Average state estimation error over an attacking period vs. attack magnitudes in case of a node misbehavior.}
 \label{fig:MSE_vs_magnitudes_misb}
\end{figure}

We assume that at time $\eta$, the node 3 is hacked and controlled by a malicious entity that injects false data to the local measurements and also does not employ the local measurement anomaly detection mechanism. Similar to the Case 1 above, we consider the FDI attack in \eqref{eq:FDI-attack} with $a_{i,t}^{3} \sim \mathcal{U}[0, \rho]$, $\forall i \in \{1,\dots,5\}$, $\forall t \geq \eta$. The difference is that the node 3 does not notify the other local centers about the local measurement anomaly at all. In this case, we expect that the misbehavior is detected over the network via the distributed trust management scheme. Assuming $\eta = 1$, we firstly present in Fig.~\ref{fig:delay_vs_mag_misb} the average detection delay versus the attack magnitude curve. Then, we present in Fig.~\ref{fig:MSE_vs_magnitudes_misb} the average state estimation performance over an attacking period of $t \in [1,50]$. We observe that via the trust management scheme and the state recovery mechanism employed after misbehavior detection, the state estimation error can be significantly reduced compared to the estimators with no recovery mechanisms.

\section{Concluding Remarks} \label{sec:conc}

We have studied the secure distributed dynamic state estimation in the wide-area smart grids. We have proposed a novel BC-based resilient mechanism in which each local center supervises a subregion and collaborates with the other local centers to perform reliable state estimation in a fully-distributed manner under adversarial settings. We have proposed to enhance the security of each part of the network against attacks and data manipulations. Particularly, we reduce the risk of manipulations in the network database and the network communication channels via cryptography and voting-consensus mechanisms. Moreover, we timely and reliably detect and eliminate the effects of measurement anomalies and misbehavior of the local control centers. Numerical studies illustrate the promising performance of the proposed mechanism.

We consider this work as a proposal for secure distributed state estimation in the wide-area smart grids, where all parts of the solution design, namely the BC mechanism, the distributed state estimation mechanism, the measurement anomaly detection scheme, the trust management scheme, and the state recovery scheme are open to improvements. Moreover, although we have specifically focused on the power networks, the proposed mechanism can be used for any linear dynamic networked control system after simple modifications. {Particularly, the following future research directions can be considered:}
\begin{itemize}
  \item The proposed BC-based mechanism introduces many advantages in terms of data security, availability, accessibility, and process transparency, etc., however, some new challenges arise such as storage redundancy over the network, higher network congestion, synchronization issues due to possible communication delays, higher computational complexity over the network, etc. Hence, the applicability of the proposed mechanism should be analyzed with proper metrics that quantify the benefits and the costs associated with the proposed mechanism and the traditional SCADA-based centralized implementation.
  \item {After early anomaly/misbehavior detection, identification of anomalies (types and causes) and the development of the corresponding mitigation/recovery strategies are needed to achieve a completely autonomous network operation.}
  \item {Secure resilient wide-area distributed dynamic state estimation needs to be studied for the nonlinear AC power system model with the extensive use of Phasor Measurement Units, as the future grid operation is expected to be more complex, dynamic, and uncertain with the broad integration of renewable and distributed energy resources \cite{Cai11,ArgonneLab16_2}.}
\end{itemize}

\appendix

\subsection{Proof of Lemma 1} \label{appendix:lemma1}

In the Kalman filter, given the measurements up to time $t-1$, we have
\begin{gather}\nonumber
\mathbf{x}_{t}^{j} \sim \mathbf{\mathcal{N}}(\hat{\mathbf{x}}_{t|t-1}^{j},\mathbf{P}_{t|t-1}^{j}),
\end{gather}
that implies
\begin{gather}\label{eq:est_err_v0}
\mathbf{x}_{t}^{j} - \hat{\mathbf{x}}_{t|t-1}^{j} \sim \mathbf{\mathcal{N}}(\mathbf{0},\mathbf{P}_{t|t-1}^{j}).
\end{gather}
Hence, similarly to \eqref{eq:est_err_v0}, we can write the following:
\begin{gather}\label{eq:est_err_v1}
\mathbf{x}_t^{\bar{\ell},j} - \mathbf{\hat{x}}_t^{\bar{\ell},j} \sim \mathbf{\mathcal{N}}(\mathbf{0},\mathbf{P}_{t|t-1}^{\bar{\ell},j}).
\end{gather}

Notice that \eqref{eq:est_err_v1} statistically characterizes the estimation error term, $\mathbf{x}_t^{\bar{\ell},j} - \mathbf{\hat{x}}_t^{\bar{\ell},j}$. Based on \eqref{eq:w_tilde_ell_j} and \eqref{eq:est_err_v1}, we then have
\begin{equation} \nonumber
{\tilde{\mathbf{w}}}_t^{\ell,j} \sim \mathbf{\mathcal{N}}(\mathbf{0}, \pmb{\Delta}_t^{\ell,j}),
\end{equation}
where
\begin{equation}\nonumber
\pmb{\Delta}_t^{\ell,j} \triangleq \mathbf{H}^{\bar{\ell},j} \mathbf{P}_{t|t-1}^{\bar{\ell},j} {\mathbf{H}}^{{\bar{\ell},j}^\mathrm{T}} + \sigma_w^2 \, \mathbf{I}_{K^{\ell,j}}.
\end{equation} \qed

\subsection{Proof of Corollary 1} \label{appendix:corr1}

Since $\chi_t^{\ell}$ is a continuous random variable, its cdf and hence the tail probability, $p_t^{\ell}$, are uniformly distributed $\mathcal{U}[0,1]$ \cite{Papoulis02}. Then, for the rest of the proof, we refer the reader to \cite[Theorem 2]{Necip19b}. The only difference in the proofs is that in \cite{Necip19b}, the tail probability is uniformly distributed in an asymptotic regime, while it is exactly uniform here. \qed

\bibliographystyle{IEEEtran}
\bibliography{det_refs,refs,Refs_SaTC17}

\end{document}